\documentclass[prb,aps,english,twocolumn,notitlepage]{revtex4-1} 
\usepackage[T1]{fontenc} \usepackage[latin1]{inputenc}
\usepackage{babel} \usepackage{txfonts}
\usepackage{epsfig,epsf,psfrag} 
\usepackage{graphicx} 
\usepackage{subfigure}
\usepackage{pslatex}
\usepackage{epic,eepic} 
\usepackage{color,pstcol}
\usepackage{pstricks} 
\usepackage{fancyhdr}  \parindent0em  
\def\ua{\uparrow}
\def\da{\downarrow}

\vspace{-10.0cm} 
\begin{document} \title{Topologically induced fractional Hall steps in the integer quantum Hall regime of $MoS_2$} 
\author{SK Firoz Islam}
\author{Colin Benjamin} \email{colin.nano@com}\affiliation{National institute of Science education \& Research, Bhubaneswar 751005, India }
\begin{abstract}

The quantum magnetotransport properties of a monolayer of molybdenum disulfide are derived using linear response theory.
Especially, the effect of  topological terms on longitudinal and Hall conductivity is analyzed.  The Hall conductivity exhibits fractional
steps in the integer quantum Hall regime. Further complete spin and valley polarization of the longitudinal conductivity is
seen in presence of these topological terms. Finally, the Shubnikov-de Hass oscillations are suppressed or enhanced
contingent on the sign of these topological terms.

 \end{abstract}

\maketitle
\section{Introduction}
Molybdenum disulfide $(MoS_2)$ is a new 2d material with possible application in nanoelectronics,
due to it's unique band structure\cite{novo}.  $MoS_2$ can be considered to some extent as a semiconductor
analog of monolayer graphene\cite{neto},
showing similar phenomena as quantum valley Hall effect\cite{park,carbotte,tahir1} while it is different from 
graphene in that spin Hall effect can be seen in addition to quantum valley Hall effect\cite{carbotte,tahir1}.\\
Monolayer $MoS_2$ is made of a layered structure with covalently bonded sulfur-molybdenum-sulfur atoms in which a single hexagonal
layer of molybdenum ($Mo$) atoms is sliced between two parallel planes of sulfur atoms, each $Mo$ atom co-ordinates with six sulfur ($S$)
atoms in prismatic fashion and each $S$ atom co-ordinates with three $Mo$ atoms\cite{xiao}. 
Similar to graphene, monolayer $MoS_2$ also consists of two valleys $K$ and $K'$ at the corners of it's
hexagonal Brillouin zone but with a direct band gap of $1.9$ eV\cite{bandgap}.\\
When few layers of $MoS_2$ are thinned down to a monolayer, the inversion symmetry is broken.
This in turn leads to a strong spin-orbit effect\cite{xiao},
in contrast to monolayer graphene. The direct band gap and the spin-valley coupling in $MoS_2$ have made it very
convenient for applications in optoelectronics\cite{photo1,photo2,photo3,photo4}, valleytronics\cite{transistor} and spintronics\cite{loss}.
Recently, several possible applications of monolayer  $MoS_2$ in valleytronics have been suggested experimentally\cite{photo3,photo4,valley2}.\\
Magnetotransport measurements have been one of the best ways to
probe electronic systems. In presence of a perpendicular magnetic field applied to the 2D system,
energy eigen states become quantized, i.e., form Landau levels. The formation of Landau levels
manifests itself through the appearance of quantum oscillation with inverse magnetic field, known as Shubnikov-de Hass
oscillation.
Another unique phenomena related to the perpendicular magnetic field is the quantization of
the Hall conductivity, i.e., $\sigma_{xy}=2(n+1) e^2/h$ with '$n$' an integer and $h$ is the Plank constant
in conventional 2D system. We use the term conventional for all non-Dirac like electronic systems.
This phenomena appears due to the conduction of fermions along the edge boundary caused by the incomplete cyclotron orbits.
The longitudinal conductivity ($\sigma_{xx}$) becomes completely zero between two consecutive Hall steps.\\
One of the most well studied non-conventional electronic system is graphene with Dirac like energy-wave vector
dispersion. Magnetotransport properties have been 
studied in graphene, both theoretically\cite{graphene1,tahir_gra2,vasilo_gra} as well as experimentally\cite{gra_exp1,gra_exp2}.
Later, same has been carried out in 
silicene\cite{tahir_sili,vasilo_sili}, a 2D silicon based hexagonal lattice similar to graphene but without spin  and valley degeneracy
in presence of gate voltage. Recently, magnetotransport measurements have been performed in $MoS_2$\cite{mos2_exp},
where SdH oscillations have been observed. Motivated by this work, in Ref.[\onlinecite{tahir_mos2}], the theoretical
study of magnetotransport propeties of $MoS_2$ was carried out. Landau levels crossing phenomena
has also been predicted in valence band of $MoS_2$\cite{fan}. Another experiment on magnetotransport measurements of
multilayer of $MoS_2$ have been also reported\cite{fan_exp}. Recently, it has been theoretically shown that
an electric field normal to $MoS_2$ can modify the band structure resulting in additional 
terms in the Hamiltonian which are quadratic in momentum\cite{reja,reja2}. These terms are also knows as
topological terms as they can affect the topological features of the system by influencing Berry curvature, Chern number and 
the $Z_2$ invariant\cite{reja2}.
Moreover, these terms have been shown to be tuned by external gate voltage which means
that magnetotransport phenomena can be manipulated by gate voltage also\cite{reja2}.
The magnetotransport properties in presence of scattering mechanism including these topological terms have not been analyzed so far
and we rectify this anomaly here.\\

In this work, we intend to present a theoretical analysis of the consequences 
of the topological terms in magnetotransport properties of $MoS_2$ by
using linear response theory. We obtain the Landau level
energy spectrum and eigen states, and the density of states for different values
of the topological parameter. Further, we study the effect of the topological parameter on SdH oscillations and the
quantum Hall conductivity. We found that SdH oscillations get 
suppressed and interestingly fractional Hall steps appear in the quantum Hall conductivity in the 
integer Hall regime. As an aside, we explore spin and valley polarization of the longitudinal conductivity and 
find them to be 100\% polarized in presence of the toplogical terms.\\
This paper is divided into four sections. After giving introduction in section (I), we derive Landau levels and 
corresponding eigen states in section (II), also discuss density of states here. In section (III),
we study longitudinal conductivity and quantum Hall conductivity. Finally, we give our conclusion in section (IV).

\section{Model Hamiltonian and Landau levels formation}
The electronic structure of Mo$S_2$ has been studied with ab initio as well as tight binding calculation\cite{xiao,reja2,ab_initio,burkard}.
We start with a simplified model of low energy effective Hamiltonian\cite{xiao}:
\begin{equation}
 H_0=\hbar v_F(\tau\sigma_x k_x+\sigma_y k_y)+\frac{\Delta}{2}\sigma_z,
\end{equation}
where $v_{F}=0.53\times 10^6$m/s is the Fermi velocity, $\tau=+(-)$ corresponding to valley $K$ ($K'$). 
{\bf k} is the 2d momentum, $\Delta$ is the direct band gap, $\sigma_i$'s are the 
Pauli matrices with $i=x,y,z$. Similar to the spin, electron in $MoS_2$ also have another degree of freedom,
called sublattice. The $\sigma$'s of the aformentioned Hamiltonian describe this sublattice parameters.\\

\subsection{ Spin-orbit interaction and topological parameter}

The removal of inversion symmetry, as mentioned earlier, generates strong spin-orbit interaction
which can be included in low energy effective Hamiltonian\cite{xiao,felix} as $H_{so}=\lambda\tau\frac{\sigma_z-1}{2}s$
with $\lambda $ being the strength of spin-orbit interaction and $s=\uparrow(\downarrow)$ describes the real spin
of the fermions.
As mentioned in introduction, the tight binding calculation based on seven band model have found additional diagonal terms
which are quadratic in $k$ but externally tunable by gate voltage in presence of magnetic field\cite{reja2}. The gate voltage
changes the on site energy of atoms and affects the band gap as well as other parameters.
Moreover, these terms can break the valley degeneracy in presence of perpendicular magnetic field\cite{reja2}.
These additional terms can be included\cite{reja2} as $H_{t}=(\alpha+\beta\sigma_z)\hbar^2k^2/(4m_0)$, where $\alpha$ is 
constant and $\beta$ is the topological parameter and $m_0$ is the free electron mass. Taking all these terms into account,
the total low energy effective Hamiltonian can now be expressed as\cite{reja2}:
\begin{equation}
 H=\hbar v_F(\tau\sigma_x k_x+\sigma_y k_y)+\frac{\Delta}{2}\sigma_z-\tau\frac{\lambda}{2}(1-\sigma_z)s
 +\frac{\hbar^2 k^2}{4m_0}(\alpha+\beta\sigma_z).
\end{equation}
Here, $\alpha$ captures  the difference between electron and hole effective masses, given by\cite{reja2}
$\alpha=m_0/m_{+}$, where $m_{\pm}=m_em_h/(m_h\pm m_e)$ with $m_{e}$, $m_{h}$ the electron and hole effective masses.
$\beta$ depends on electron/hole effective masses as well as on the band gap and spin-orbit interaction strength, given by\cite{reja2}
$\beta=m_0/m_{-}-4m_0v_F^2/(\Delta-\lambda)$.\\

\subsection{ Inclusion of magnetic field:}

When  an uniform perpendicular magnetic field is applied to  monolayer of $MoS_2$, energy of conduction and valence band becomes
quantized, i,e., Landau levels are formed without spin and valley degeneracy.
\begin{figure}
\begin{center}
\includegraphics[width=.45\textwidth,height=40mm]{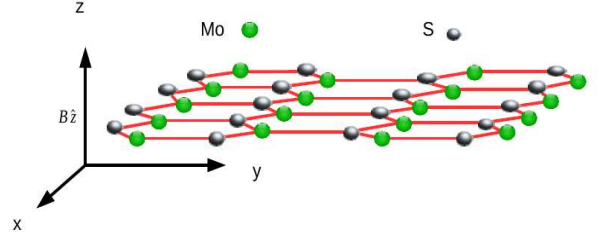}
\caption{A top view of a monolayer $MoS_2$.}
\end{center}
\end{figure}
Solving the Hamiltonian in presence of the magnetic field, we obtain Landau levels and corresponding
eigen states. The magnetic field is included via Landau-Peierls substitution: ${\bf p}\rightarrow {\bf p+eA}$ in single electron Hamiltonian
of monolayer $MoS_2$ in the x-y plane as
\begin{equation}
  H= v_{F}{\bf \sigma_{\tau}}.{ \Pi}+\frac{\Delta}{2}\sigma_z-\tau\frac{\lambda}{2}(1-\sigma_z)s
 +\frac{\Pi^2}{4m_0}(\alpha+\beta\sigma_z),
\end{equation}
where $\Pi={\bf (p+eA)}$ is 2d momentum operator with e the electronic charge, $\sigma_\tau=(\tau\sigma_x,\sigma_y)$
and ${\bf A}$ is the magnetic vector potential,
choosing Landau gauge ${\bf A}=(0,xB,0)$ describes the magnetic field ${\bf B}=B\hat{z}$. \\
We introduce dimensionless harmonic oscillator operators $a=\frac{1}{\sqrt{2}}(P_x-iX)$ and $a^{\dagger}=\frac{1}{\sqrt{2}}(P_x+iX)$,
where $X=(x+x_0)/l_c$ with the origin of the cyclotron orbit at $x=-x_0$, where $x_0=k_y l_c^2$ and $l_c=\sqrt{\hbar/eB}$-the magnetic length.
Also, $P_x=-i\partial/\partial(x/l_c)$-the dimensionless momentum operator. Then the above Hamiltonian for K-valley ($\tau=+1$) takes the form as
\begin{equation}\label{eqn2}
H= \left[\begin{array}[c]{c c}
 \frac{\Delta}{2}+E_{2d}(a^{\dagger}a+\frac{1}{2})\gamma_{+} & \sqrt{2}E_g a\\
   \sqrt{2}E_g a^{\dagger} &-\frac{\Delta}{2}+\lambda s+E_{2d}(a^{\dagger}a+\frac{1}{2})\gamma_{-}
        \end{array}\right],
\end{equation}
where $E_{2d}=\hbar eB/m_0$ and $E_g=\hbar v_F/l_c$, the two energy scales with $l_c=\sqrt{\hbar/eB}$-magnetic length,
and, $\gamma_{\pm}=(\alpha\pm\beta)$. It should be noted that two energy scales arise in the solution of Hamiltonian(\ref{eqn2}).
This is in contrast to the case where the topological parameters are absent\cite{tahir_mos2}. The energy scale $E_g$ describes
Dirac like dispersion while $E_{2d}$ describes conventional non-Dirac like dispersion. So the influence of the 
topological parameter is to add conventional 2D electronic features into a Dirac-like system.\\
To diagonalize the above Hamiltonian, we choose the spinor
\begin{equation}
 \Phi(X)=\frac{1}{\sqrt{2}}\sum_{j}\left[\begin{array}[c]{c} c_j^{+} \\c_{j}^{-}\end{array}\right]\phi_{j}(X)
\end{equation}
as the basis, where $\phi_j(X)$ is harmonic oscillator wave function. Here, $j=0,1,2,3..$. $c_{j}^{\pm}$
are the unknown coefficients of upper(+) and lower(-) components spinor.
Using this wave function in the eigen value
equation $H\Phi=E\Phi$ and multiplying the same harmonic oscillator wave function with different index $l$, $\phi_l(x)$ from left side,
we obtain the following set of coupled equations, after
integrating and using orthogonal properties of Hermite polynomials we get 
\begin{equation}\label{1st}
 [E_{2d}(l+\frac{1}{2})(\alpha+\beta)+\frac{\Delta}{2}-E]c_{l}^{+}+E_g\sqrt{2(l+1)}c_{l+1}^{-}=0
\end{equation}
and
\begin{equation}\label{2nd}
 E_g\sqrt{2l}c_{l-1}^{+}+[E_{2d}(l+\frac{1}{2})(\alpha-\beta)-\frac{\Delta}{2}+\lambda s-E]c_{l}^{-}=0.
\end{equation}
In Eq.(\ref{1st}), we make transformation $(l+1)\rightarrow n$, a new index while in Eq.(\ref{2nd}) we keep $l\rightarrow n$.
This is the usual way for solving eigen value problem for 2D electronic systems\cite{vasilo_rashba}.\\
Finally, we have
\begin{equation}\label{3rd}
 [E_{2d}(n-\frac{1}{2})(\alpha+\beta)+\frac{\Delta}{2}-E]c_{n-1}^{+}+E_g\sqrt{2n}c_{n}^{-}=0
\end{equation}
and
\begin{equation}\label{4th}
 E_g\sqrt{2n}c_{n-1}^{+}+[E_{2d}(n+\frac{1}{2})(\alpha-\beta)-\frac{\Delta}{2}+\lambda s-E]c_{n}^{-}=0,
\end{equation}
here $n=1,2,3,4...$ are the Landau level index. Similar coupled equations can be obtained for $K'$-valley too. Solving Eqs. (\ref{3rd}-\ref{4th}), 
we get Landau levels as:
\begin{eqnarray}\label{ll}
 E_{\xi}&=&\frac{\lambda \tau s}{2}+(\alpha n-\tau\frac{\beta}{2})E_{2d}+\nonumber\\&p&
 \sqrt{2nE_g^2+[\frac{\Delta-\lambda \tau s}{2}+E_{2d}(\beta n-\tau\frac{\alpha}{2})]^2}.
\end{eqnarray}
Here, $\xi=\{k_y,\zeta\}$ with $\zeta=\{n,s,\tau,p\}$, where $p=+(-)$ stands for conduction (valence) band. Note that in absence of the topological
parameter, energy levels of $K_{\ua(\da)}$ is equivalent to $K'_{\da(\ua)}$ but this symmetry is now broken.
The magnetic field dependency of Landau levels can be shown as
\begin{eqnarray}
  E_{\xi}(B)&=&\frac{\lambda \tau s}{2}+(\alpha n-\tau\frac{\beta}{2})\frac{\hbar eB}{m_0}+\nonumber\\&p&
 \sqrt{2n\frac{\hbar^2 v_F^2 eB}{\hbar}+\big[\frac{\Delta-\lambda \tau s}{2}+\frac{\hbar eB}{m_0}(\beta n-\tau\frac{\alpha}{2})\big]^2}.
\end{eqnarray}
The ground state energy in K-valley can be obtained from Eq.(\ref{2nd}) as $ E_{0,+,s}=\frac{E_{2d}}{2}(\alpha-\beta)-\frac{\Delta}{2}+\lambda s
$. Similarly, for K'-valley which is independent of spin-orbit interaction $ E_{0,-}=\frac{E_{2d}}{2}(\alpha+\beta)+\frac{\Delta}{2}$.
Note that ground state energy is negative in K-valley while it is positive in K'-valley, as $\Delta/2$
is much higher than $E_{2d}$ and $\lambda$.\\
The corresponding wave functions in both valleys are
\begin{equation}
 \Phi_{\xi}({x,y})=\frac{e^{ik_yy}}{\sqrt{L_y}}\left[\begin{array}[c]{c}\tau A_{n,s}^{\tau,p}\phi_{n-1}\big[\frac{x+x_0}{l_c}\big]\\
                                                       B_{n,s}^{\tau,p}\phi_{n}\big[\frac{x+x_0}{l_c}\big]
                                                      \end{array}\right]
\end{equation}
with
\begin{equation}\label{13}
 A_{n,s}^{\tau,p}=\frac{E_{g}\sqrt{2n}}{\sqrt{2nE_g^2+[E_{2d}(n-\frac{\tau}{2})(\alpha+\beta)+\frac{\Delta}{2}-E_{\zeta}]^2}}
\end{equation}
and
\begin{equation}\label{14}
 B_{n,s}^{\tau,p}=\frac{E_{\xi}-E_{2d}(n-\frac{\tau}{2})(\alpha+\beta)-
 \frac{\Delta}{2}}{\sqrt{2nE_g^2+[E_{2d}(n-\frac{\tau}{2})(\alpha+\beta)+\frac{\Delta}{2}-E_{\zeta}]^2}}.
\end{equation}
Here, $e^{ik_yy}/\sqrt{L_y}$ is the normalization factor.
$\phi_n(X)=\frac{1}{\sqrt{2^n n!l_c\sqrt{\pi}}}e^{-X^2/2}H_n(X)$ is the harmonic oscillator wave function with $H_n(X)$
the Hermite polynomial of order n. The harmonic oscillator wave functions $\phi_{n-1}(x)$ and $\phi_{n}(x)$ are interchanged in K'-valley.
The ground state wave function in K-valley is
\begin{equation}
 \Phi_{k_y,0,+}(x,y)=\frac{e^{ik_yy}}{\sqrt{L_y}}\left[\begin{array}[c]{c}0\\\phi_{0}(x)
                   \end{array}\right]
\end{equation}
and in K'-valley is
\begin{equation}
 \Phi_{k_y,0,-}(x,y)=\frac{e^{ik_yy}}{\sqrt{L_y}}\left[\begin{array}[c]{c}\phi_0(x)\\ 0
                   \end{array}\right].
\end{equation}
 \begin{figure*}
\centering
   \subfigure[]{\includegraphics[width=.48\textwidth,height=28em]{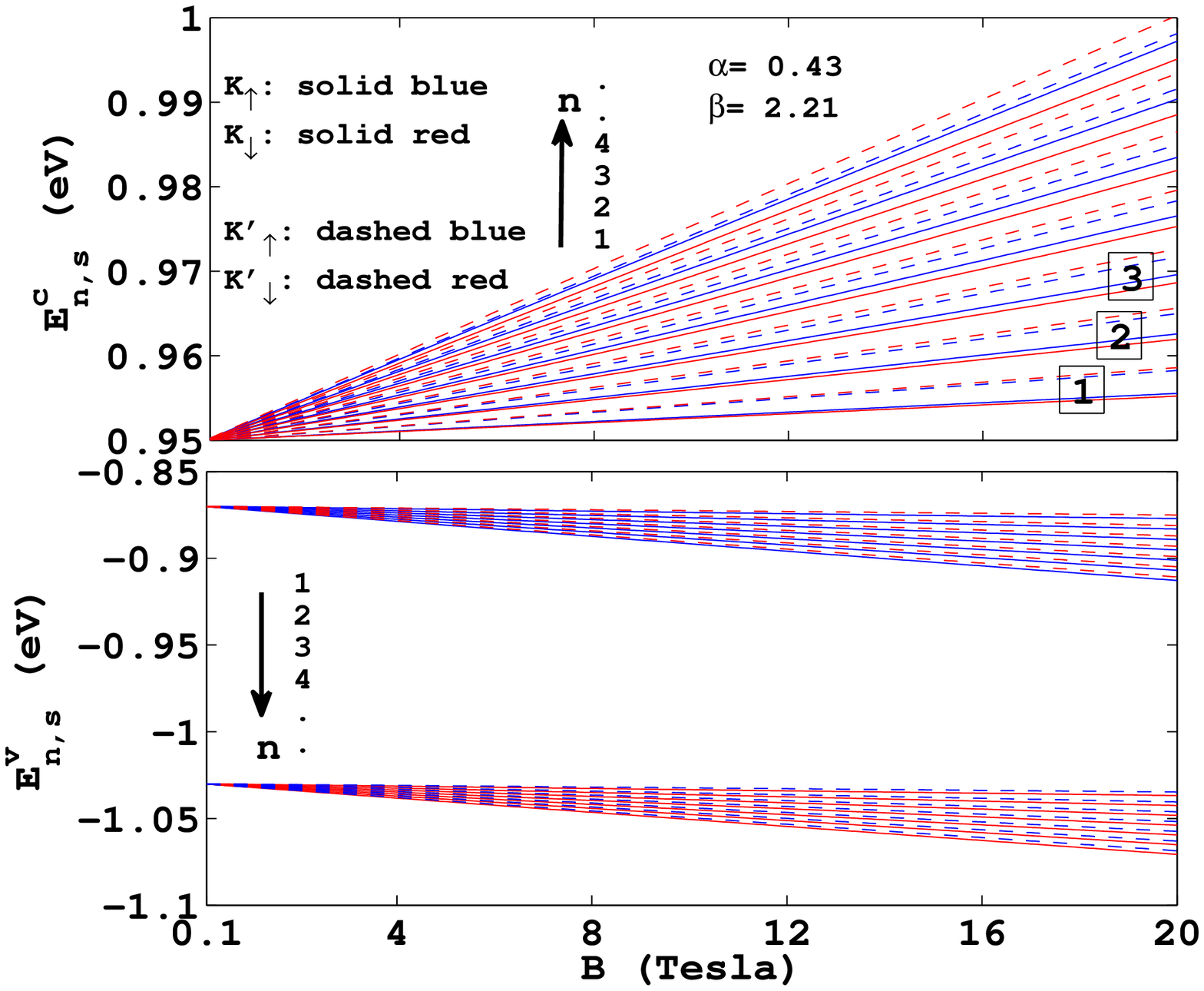}}
   \subfigure[]{\includegraphics[width=.48\textwidth,height=28em]{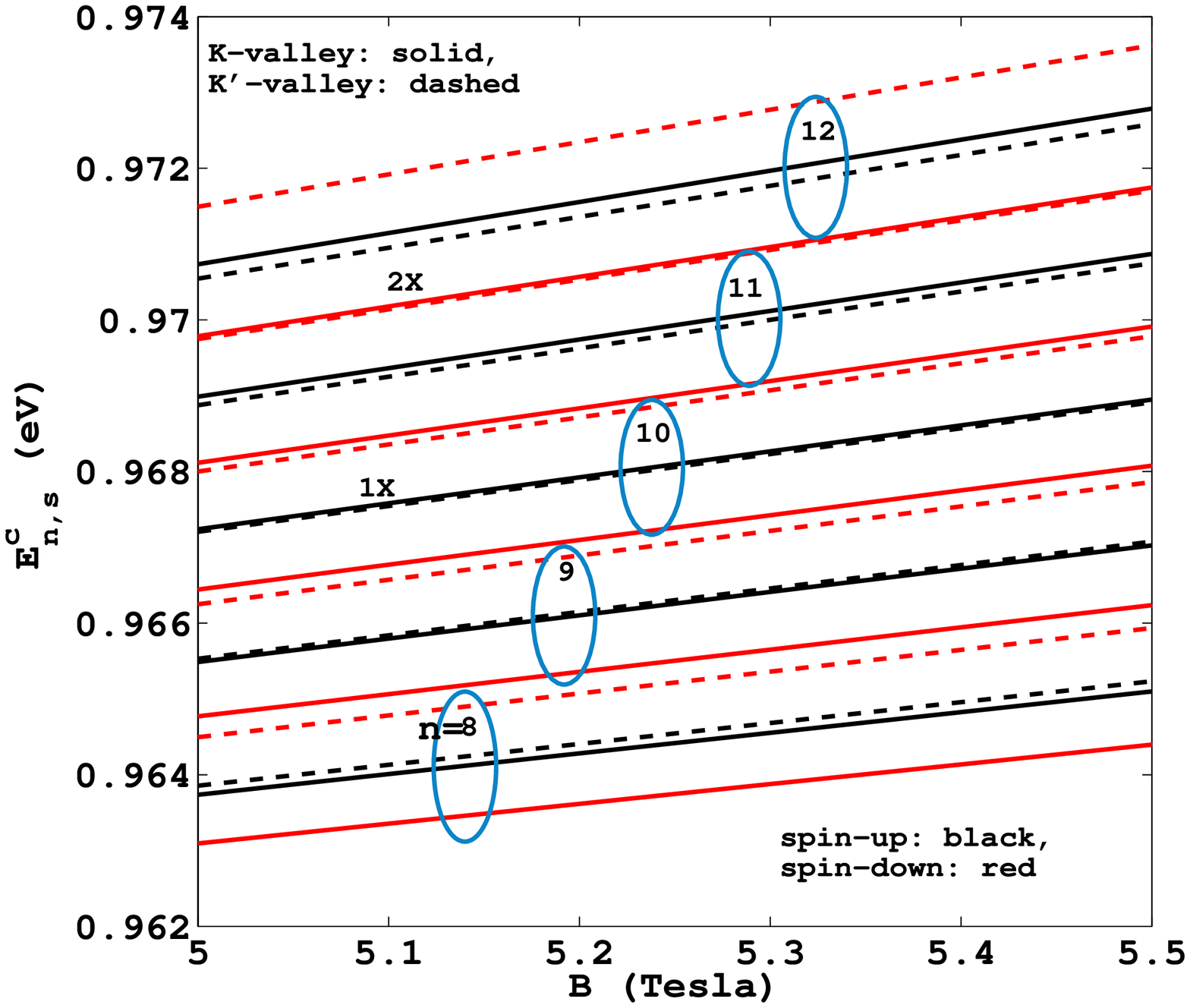}}
    \caption{Landau levels in both valley and in conduction (upper panel) and valence (lower panel) band in left figure.
    The right figure shows the zoomed part of few Landau levels between $B=5$ to $5.5$ T. Parameters: $v_{F}=0.53 \times 10^5$ m/s, direct band gap $\Delta=1.9$ eV and spin-orbit interaction strength $\lambda=0.08$ eV.  1X and 2X mark the places of vanishing Landau level spin splitting, see text in Section II B.}
   \label{Fig2}
 \end{figure*}
 We plot the spin and valley dependent Landau levels in Figure. (\ref{Fig2}a).
 we see that Landau levels in the conduction band linearly increases with magnetic field, while in the valence band
 they linearly decrease with the magnetic field.\\
 Landau levels have two magnetic field dependent energy scales, one is topologically induced $E_{2d}\propto B$ and the Dirac nature induced 
 $E_{g}\propto\sqrt{B}$, which compete with each other.\\ 
 Now we only discuss conduction band Landau levels, as we consider Fermi level lies in the conduction band.
 Landau levels are always well separated in valley space but not so in spin space.
 In fact for smaller indices, the spin splitting is unobservable. In absence of the topological parameter, spin splitting
 would be same in both valleys  but opposite i.e., energy levels corresponding to $K_{\uparrow}$ ($K_{\downarrow}$) is equivalent to
 $K^{\prime}_{\downarrow}$ ($K^{\prime}_{\uparrow}$), thus each Landau level is still doubly degenerate. The topological 
 parameter is intrinsically associated with the Landau level index as shown in Eq.(\ref{ll}), see the second term under the square root
 the topological factor $\beta$ is multiplied with $n$ which is in complete contrast to the valley dependent Zeeman effect as treated
 in Ref.[\onlinecite{tahir_mos2}].
 At strong magnetic field, the effect of the topological parameter is expected to be dominant over Dirac kinetic energy terms.
 In the valence band (lower panel of Figure (\ref{Fig2}a), spin splitting is very strong in both valleys. 
 We also see a fascinating phenomena of vanishing Landau level spin splitting in valley space,
 see for example the Landau level index $n=10$ where $E_{10}(K_{\downarrow})\cong E_{10}(K'_{\downarrow})$,
 indicated by 1X. For higher Landau levels again, we see a similar thing but now not in the same level
 but between adjacent Landau levels, see the point marked 2X in figure
 ({\ref{Fig2}b) i.e.,$E_{11}(K_{\downarrow})\cong E_{12}(K'_{\downarrow})$.
 To conclude, we see that intra and inter valley and spin dependent Landau level gaps may disappear
 as a consequence of topological parameter. This has major consequence for SdH oscillations and quantum Hall conductivity as
 we describe in the next section, but before that we now concentrate on the density of states.\\
 
 \subsection{ Landau levels in presence of Zeeman term}
 Note that so far we have not included Zeeman effects as treated in Ref.[\onlinecite{tahir_mos2}].
 If we consider the Zeeman effect too, then there will be an 
 additional term in total Hamiltonian $H'=H+H_{z}$, where $H_{z}=-\vec{\mu}.\vec{B}$. Here, $\vec{\mu}$ is the magnetic moment of electron,
 which can be written as $\vec{\mu}=g_{s}\mu_{_B}\vec{S}/2$. Then Zeeman term reduces to $H_{z}=-g_{s}\mu_{_B}\sigma_{z}B/2$. Now
 following the same procedure we get Landau levels in presence of Zeeman term as
 \begin{eqnarray}\label{zeeman}
 E_{\xi} &=&\frac{\lambda \tau s}{2}+(\alpha n-\tau\frac{\beta}{2})E_{2d}\nonumber\\&+&
 \sqrt{2nE_g^2+[\frac{\Delta-\lambda \tau s}{2}-E_{z}+E_{2d}(\beta n-\tau\frac{\alpha}{2})]^2}.
\end{eqnarray}
Here, Zeeman energy $E_{z}=g_s\mu_{_B}B/2=g_{s}E_{2d}/4$. We have checked that topologically induced energy correction is
$3.2$ meV at $B=0.5$ T for a typical Landau level $n=25$,
on the other hand Zeeman energy is $0.057$ meV which is too small and can be ignored as compared to the spin-orbit
as well as the terms containing topological parameters $\alpha$ and $\beta$.
Zeeman terms, baring Ref.~[\onlinecite{tahir_mos2}] which of course considers magnetically doped $MoS_2$, are usually ignored. See for example Refs.[\onlinecite{fan}]
and [\onlinecite{felix}] where Zeeman terms are ignored for aforesaid reasons.

Now, we shall point out few distinct 
features of the above Landau levels (Eq.\ref{ll}) which are in contrast to the case
of without topological terms but in presence of spin and valley dependent
Zeeman term as considered in Ref. [\onlinecite{tahir_mos2}]. The spin splitting Landau levels ($E_{n,\uparrow,\tau,p}
-E_{n,\downarrow,\tau,p}$) are valley dependent and tunable by varying magnetic field and controlling topological parameters
via gate voltage. On the other hand in Ref.[\onlinecite{tahir_mos2}], though 
a spin dependent Zeeman term causes valley dependent spin splitting Landau levels, but the scope of tuning
the spin splitting is limited to magnetic field only. The topological parameters are 
intrinsically associated with Landau level index `$n$' which suggests that spin splitting will depend on `$n$' strongly.
On the other hand in Ref. [\onlinecite{tahir_mos2}], as spin dependent Zeeman term
does not depend on `$n$' the spin splitting Landau levels depend on `$n$' weakly.
 
 \subsection{ Density of states}
  \begin{figure*}
    \centering
   \subfigure[Topological parameters: $\alpha=0.43$ and $\beta=2.21$]{\includegraphics[width=.47\textwidth,height=20em]{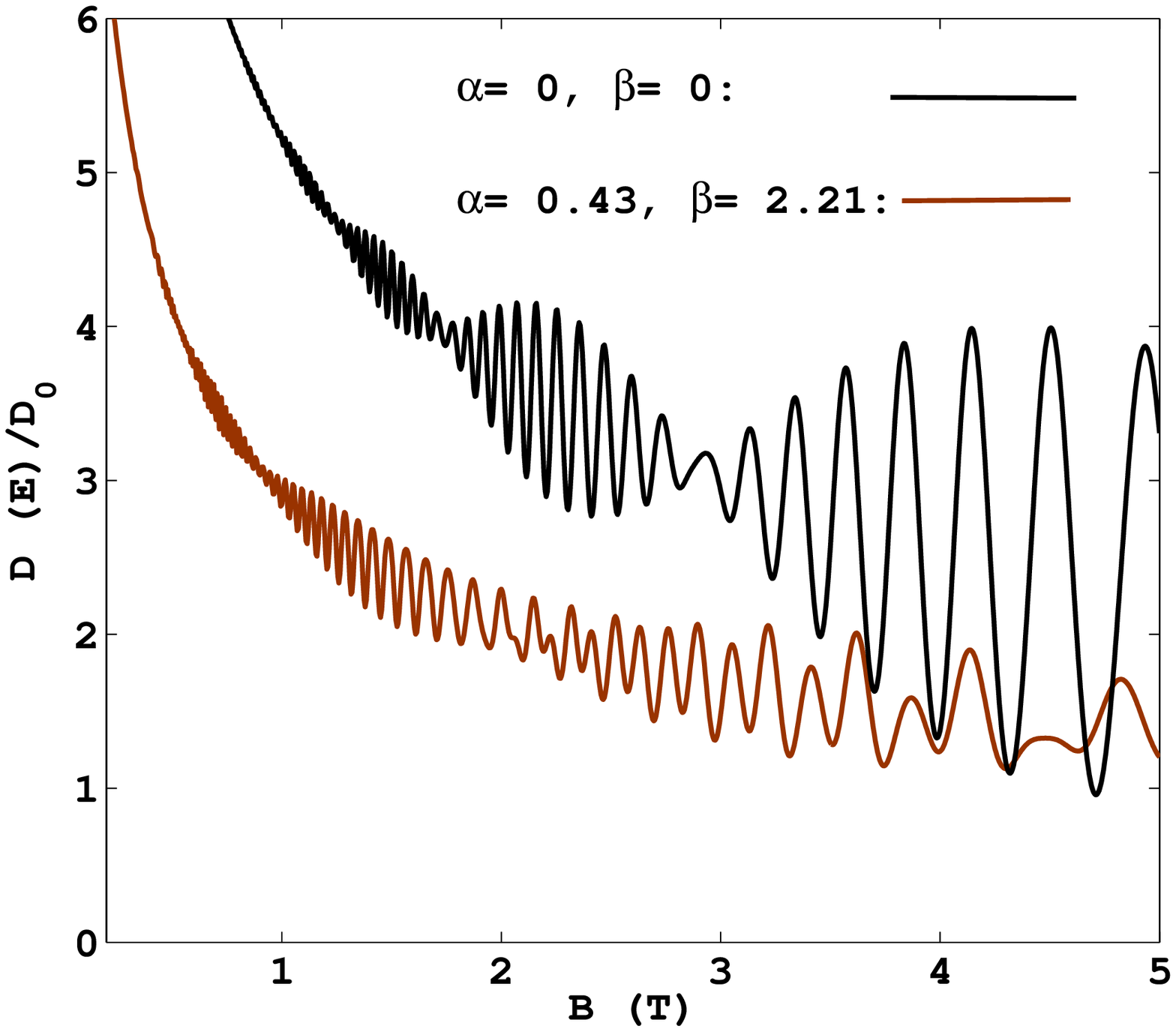}}
   \subfigure[Topological parameters: $\alpha=-0.01$ and $\beta=-1.54$]{\includegraphics[width=.47\textwidth,height=20em]{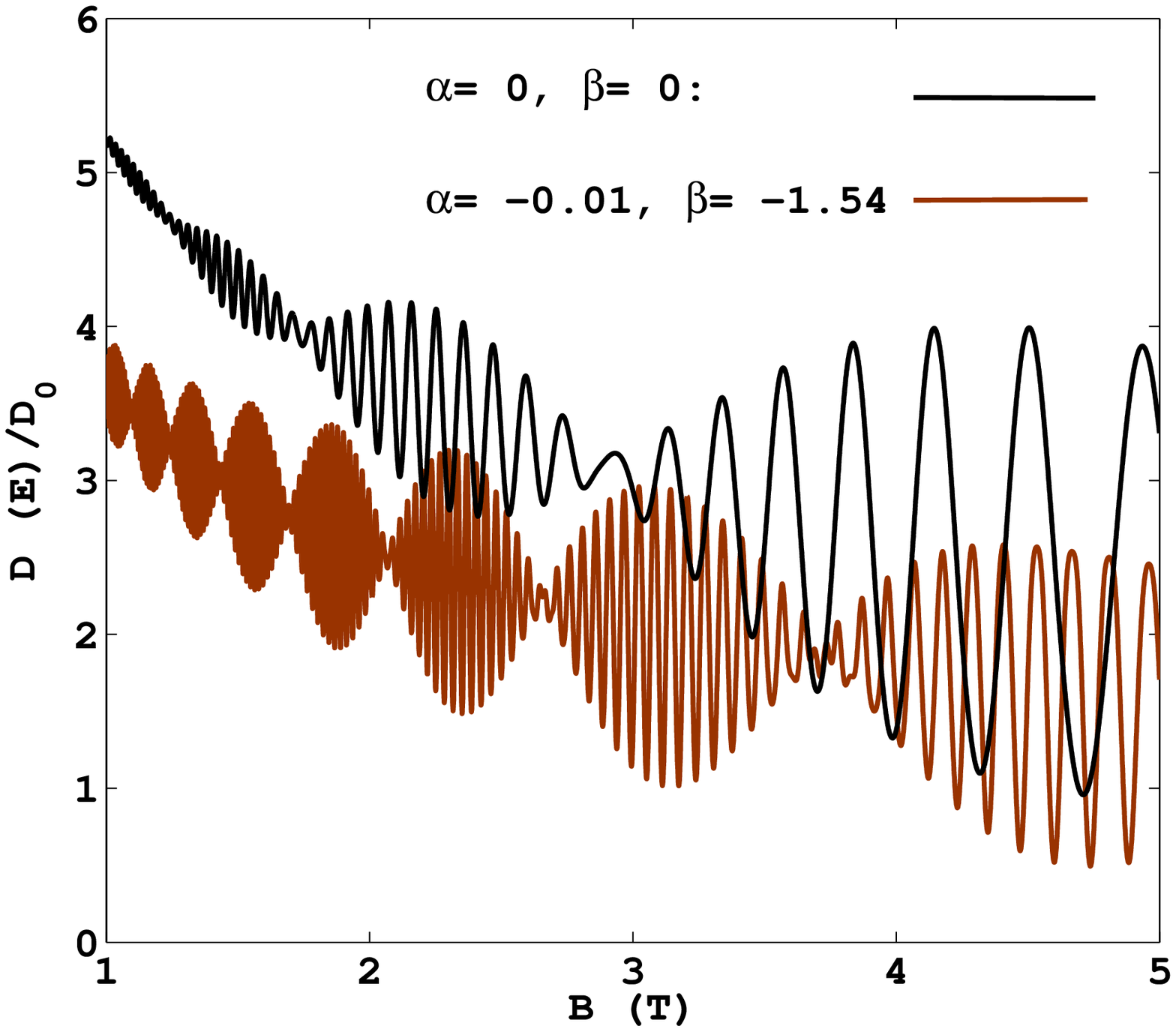}}
  \centering
   \caption{Plots of density of states in conduction band versus magnetic field
            without and with topological parameters at Fermi energy $E_{f}=0.96$ eV. Two different values of
            topological parameters ($\alpha$ and $\beta$) are used, which can be obtained by suitably choosing applied gate voltage.
            Here, number of beating nodes as well as frequencies of SdH oscillation are shown to be strongly modified.}
   \label{Fig3}
 \end{figure*}
 \begin{figure}
\begin{center}
\includegraphics[width=.45\textwidth,height=60mm]{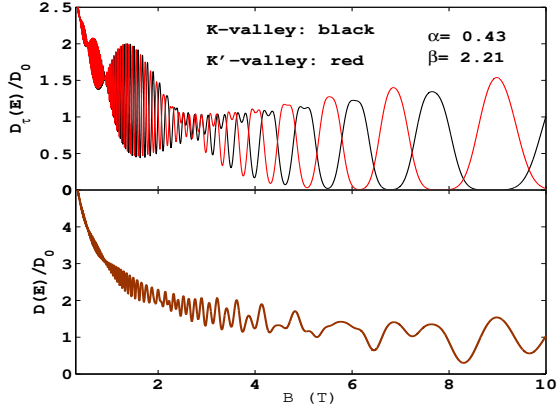}
\caption{Density of states plots in each valley for different values of topological parameters are shown.
Positive values of topological parameters cause complete valley separation at $B>5$T as shown in top panel.
The lower panel shows weak separation in valley space for negative values topological parameters in a
wide range of magnetic field.}
\label{Fig4}
\end{center}
\end{figure}
 The density of states (DOS) in presence of perpendicular magnetic field would be a series of delta function because
 of the discrete energy spectrum expressed as
 \begin{equation}
D(E)=g_sg_v\sum_{k_y,\zeta}\delta[E-E_{k_y,\zeta}],
\end{equation}
where, $g_s$ and $g_v$ are spin and valley degeneracy, respectively.\\
The summation over $k_y$ can be evaluated from the fact that origin of the cyclotron orbit is limited by the system dimensions
i.e., $0\le \mid x_0\mid \le L_x$, or $0\le k_y\le L_x/l_c^2$, thus
\begin{equation}
 \sum_{k_y}\rightarrow \frac{L_y}{2\pi}\int_0^{L_x/l_c^2}dk_y=\frac{\Omega}{2\pi l_c^2},
 \end{equation}
 with $\Omega=L_xL_y$. Here, the factor $L_y/(2\pi)$ appears from the 
 boundary condition. Now the DOS per unit area can be expressed as  
 \begin{equation}
 D(E)=C_0\sum_{\zeta}\delta[E-E_{\zeta}],
 \end{equation}
where $C_{0}=g_sg_v/(2\pi l_c^2)$ with $g_s=g_v=1$ as there is no spin and valley degeneracy.
To plot DOS, we shall use the Gaussian distribution of the delta function as
\begin{equation}
D(E)=D_{0}\sum_{\zeta}\exp[{-\frac{(E-E_{\zeta})^2}{2\Gamma^2}}],
\end{equation}
where $D_{0}=C_{0}/\sqrt{2\pi}\Gamma$ with the width of the Gaussian distribution
$\Gamma=0.1\sqrt{B}$ meV\cite{broadening}. DOS become oscillatory in 
presence of magnetic field due to the quantization of energy spectrum, see Figure (\ref{Fig3}).
In absence of the topological parameter, both K and K'-valleys produce beating pattern in DOS oscillations
with beating nodes at the same location. This is expected as spin splitting in both valleys is same but opposite i.e., 
$E_{K}^{\uparrow(\downarrow)}\cong E_{K'}^{\downarrow(\uparrow)}$.
The total DOS without topological parameter is shown in black in Figure(\ref{Fig3}a and b).
Beating pattern is caused by the superposition of two closely separated frequencies of DOS for spin-up and down branches.
The effects of inclusion of topological terms are shown in brown in Figures(\ref{Fig3}a) and (\ref{Fig3}b). Figure(\ref{Fig3}a) shows
that in the range of magnetic field $B>1$ T, beating pattern disappears and suppression of SdH oscillations occurs for $\alpha=0.43$
and $\beta=2.21$. The Figure(\ref{Fig3}b) shows that SdH oscillations in DOS are pronounced with increased frequency and higher number
of beating nodes for $\alpha=-0.01$ and $\beta=-1.54$.\\
To understand these behavior of DOS in presence of topological parameters, we plot DOS in each valley separately
in Figure(\ref{Fig4}). Here we see that there is a definite phase difference between two valleys
which resulted in suppression of the total DOS oscillation (lower panel) for $\alpha=0.49$
and $\beta=2.21$ as shown in the upper panel of Figure(\ref{Fig3}a). But this is not followed when $\beta=-1.54$ in lower panel
of Figure(\ref{Fig4}), where the phase difference between two valleys is small, and sustains DOS oscillation in total DOS
with spin-split induced beating pattern. Note that suppression of SdH oscillations is also shown in Ref. [\onlinecite{tahir_mos2}]
in presence of spin and valley dependent Zeeman terms, but the new features what we see here are the 
disappearance of beating pattern above $B>1$ as shown in Figure(\ref{Fig3}a) and enhancement of the SdH oscillations
as well as beating nodes and frequency as shown in Figure(\ref{Fig3}b).

\section{Electrical conductivity}
 At low temperature regime, there are mainly two kind of mechanisms by which electronic conduction takes place.
 One is due to the scattering of cyclotron orbits from localized charged impurities-
 the collisional contribution to the conductivity. The other contribution is the diffusive 
 conductivity which depends on the drift velocity of the electron\cite{peeters_92}.
 Since drift velocity $v_{i}=\hbar^{-1}\partial E/\partial k_{i}=0$ in our case with $i={x,y}$,
 therefore diffusive contribution to the conductivity vanishes.
 To calculate different components of conductivity tensor (Hall conductivity-$\sigma_{xy}$
 and the longitudinal conductivity -$\sigma_{xx/yy}$), we shall use linear response theory
 modified in Ref.[\onlinecite{theory}].

\subsection{ Longitudinal conductivity}
Here, we assume that electrons are scattered 
elastically by randomly distributed charged impurities. This types of scattering is important at low temperature.
The expression for longitudinal conductivity is given by\cite{theory}
\begin{equation}
 \sigma_{xx}=\frac{\beta_{_T}e^2}{2\Omega }\sum_{\xi,\xi'}f_{\xi}(1-f_{\xi'})W_{\xi,\xi'}(x_{\xi}-x_{\xi'})^2.
\end{equation}
 \begin{figure*}
\centering
   \subfigure[black figure: Topological parameters $\alpha=\beta=0$; red figure: 
   Topological parameters $\alpha=0.43$ and $\beta=2.21$. Apart from the suppression of SdH 
   oscillation, topological parameters also influence the location of beating nodes.]{\includegraphics[width=.45\textwidth,height=20em]{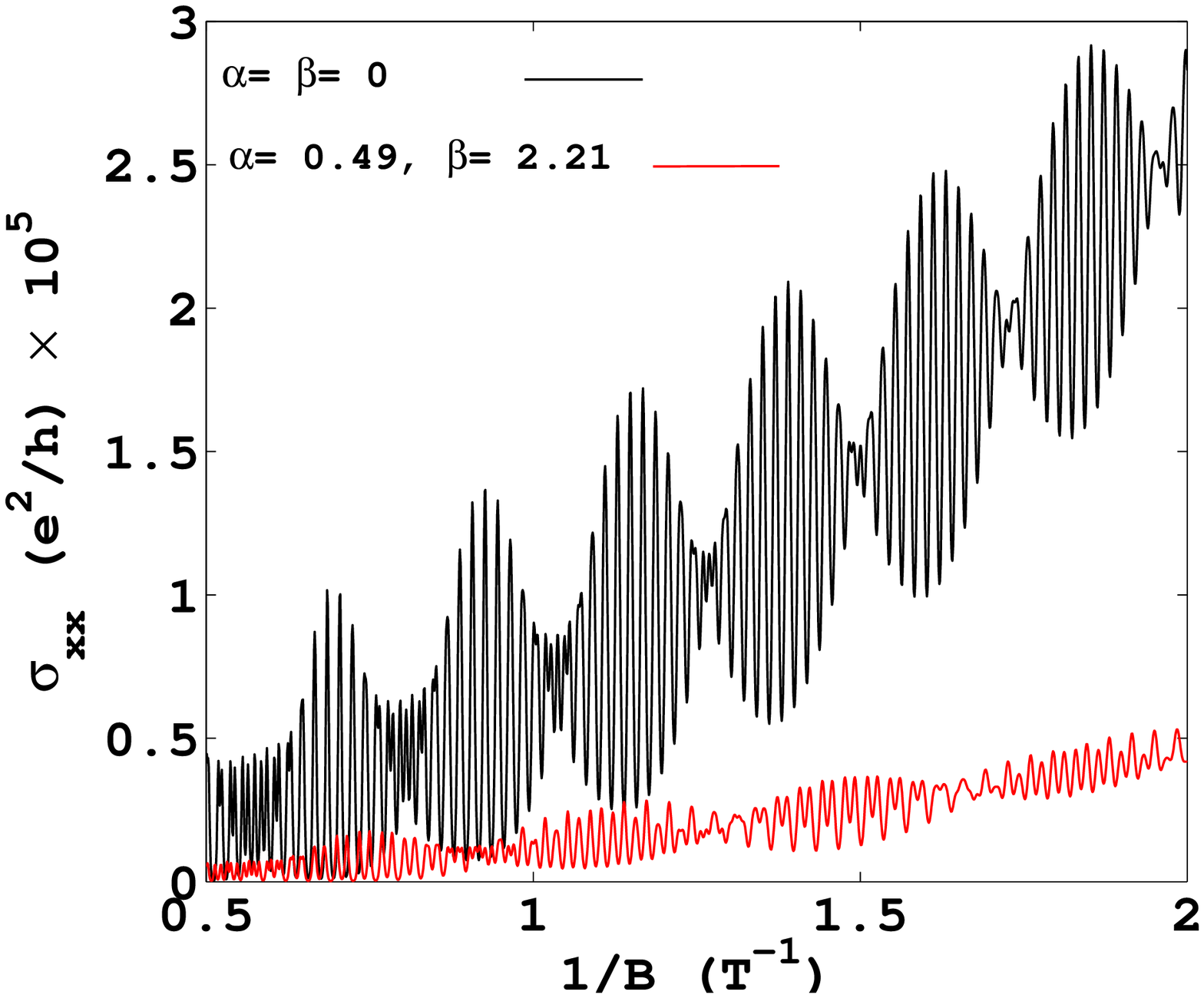}}
   \subfigure[Topological parameters are used as $\alpha=-0.01$ and $\beta=-1.54$. Here, SdH oscillation 
   frequency as well as number of beat nodes both are increased. However, amplitude of SdH oscillation 
   gets enhanced instead of getting damped as in the case of Fig.(\ref{Fig5}a)]{\includegraphics[width=.45\textwidth,height=20em]{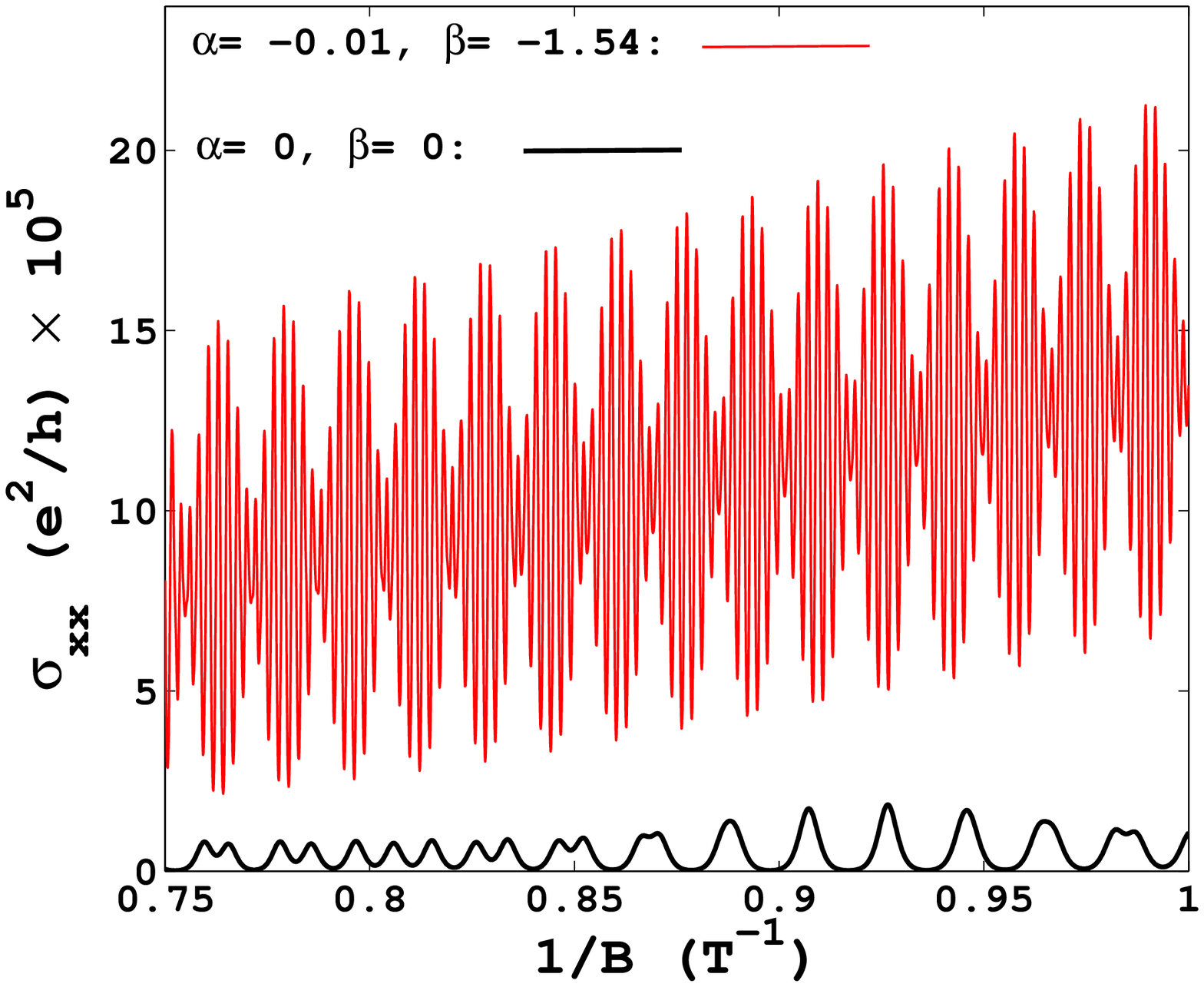}}
   \centering
   \caption{Plots of the longitudinal conductivity versus inverse magnetic field.}
   \label{Fig5}
 \end{figure*}
 
  \begin{figure*}
  \subfigure[spin polarization vs magnetic field]{\includegraphics[width=.45\textwidth,height=20em]{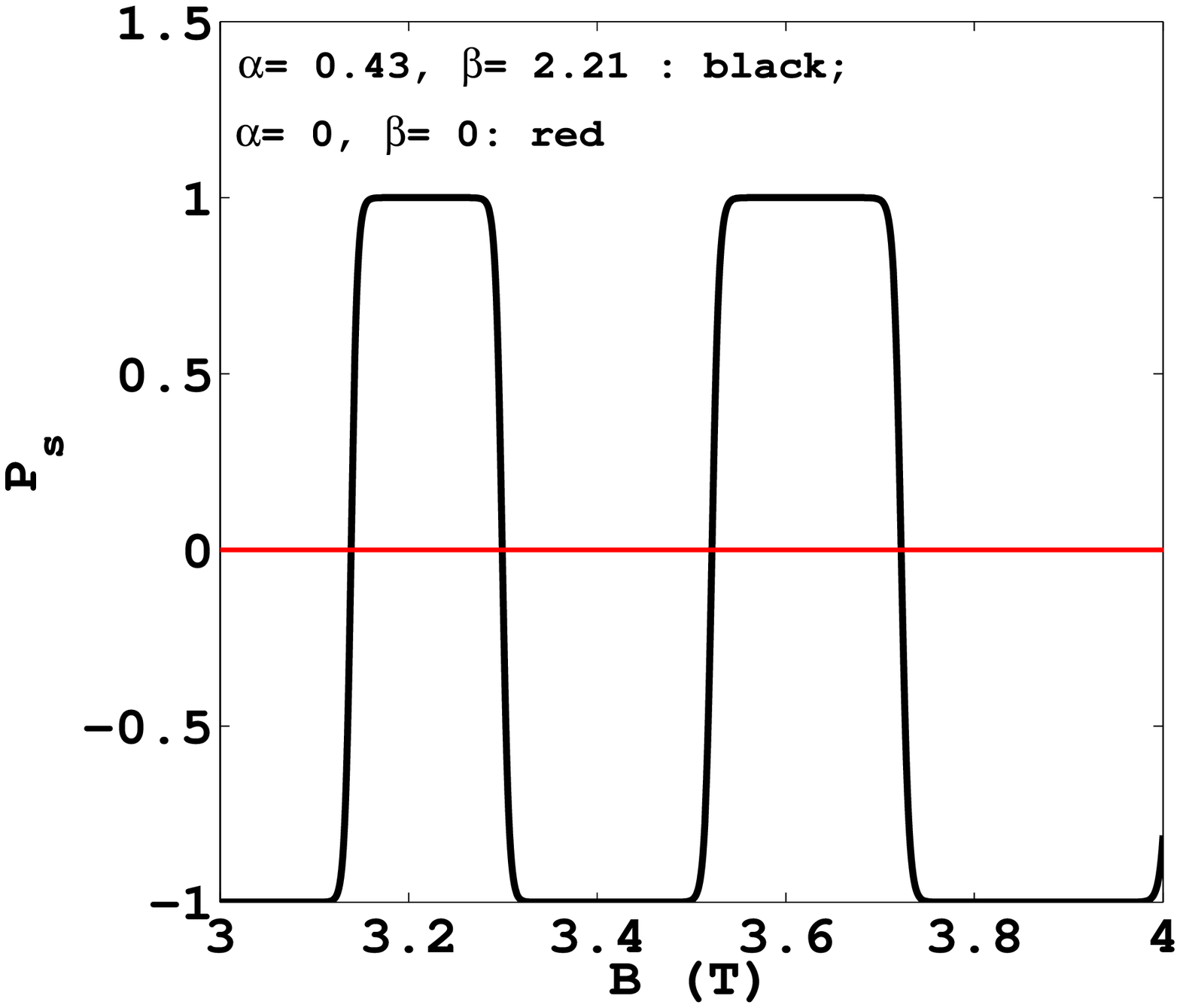}}
   \centering
   \subfigure[valley polarization vs magnetic field]{\includegraphics[width=.45\textwidth,height=20em]{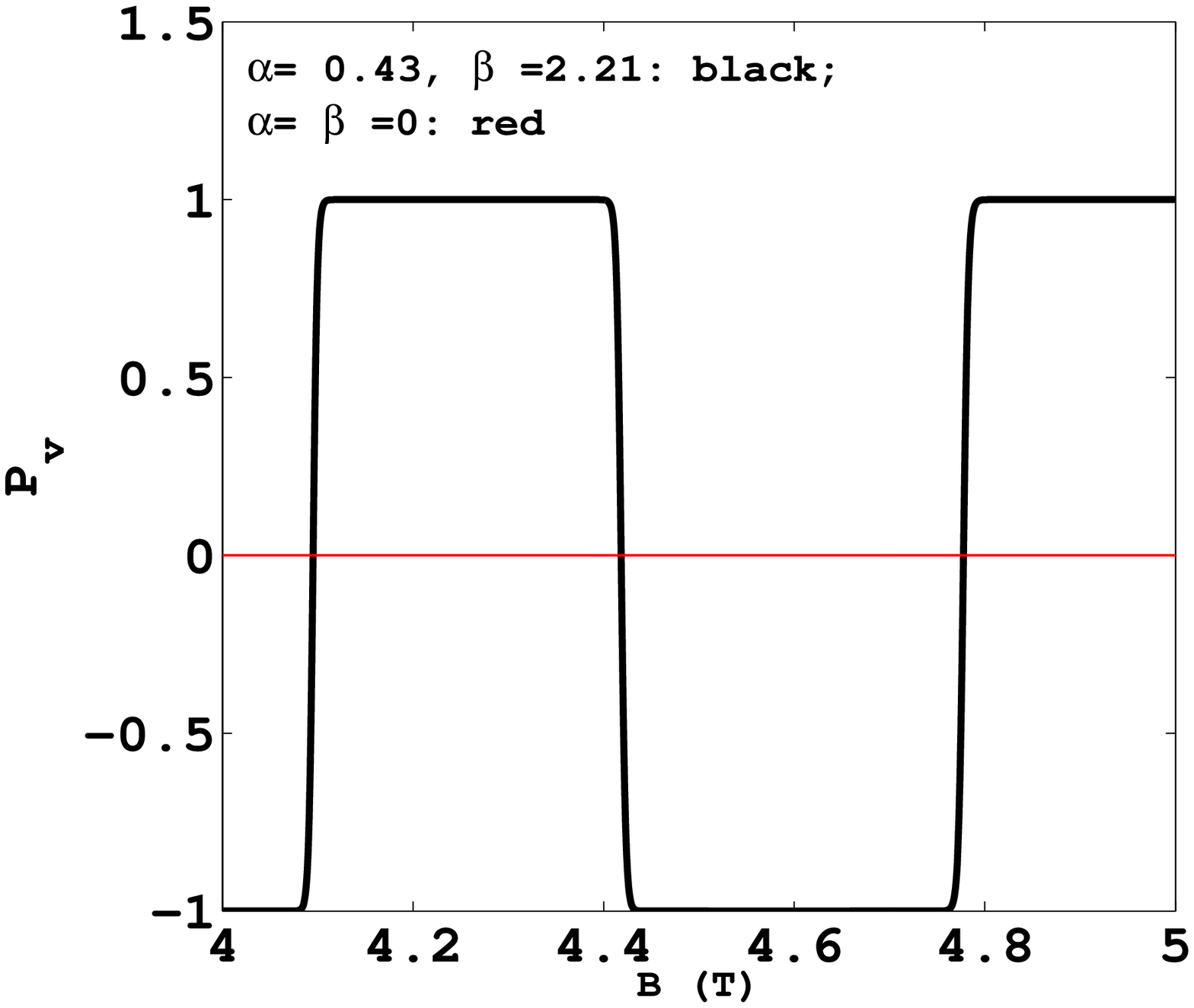}}
   \centering
   \caption{Plots of the spin/valley polarization in longitudinal conductivity versus magnetic field. A complete spin/valley polarization is
   obtained for a wide range of magnetic field but changes sign alternatively. Beat node may arise in spin/valley polarization,
   as found in valley polarization at $1/B=1.35 T^{-1}$ due to spin splitting in each valley. Without topological parameters,
   spin/valley polarization disappear.}
   \label{Fig6}
 \end{figure*}
Here, $\beta_{_T}=(k_BT)^{-1}$, $f_{\xi}=[1+\exp\{\beta_{T}(E_{\xi}-E_{F})\}]^{-1}$ is the Fermi-Dirac distribution function.
$x_{\xi}=<\xi\mid x \mid \xi>$, is the  expectation value of the x-component of the position operator when electron is in state  $\mid \xi>$. It can be
easily shown that $x_{\xi}=x_0=k_yl_c^2$ i,e; $(x_{\xi}-x_{\xi'})^2=(q_yl_c^2)^2$ with $k_y'-k_y=q_y$.
The scattering rate is given by\cite{theory}
\begin{equation}
 W_{\xi,\xi'}=\frac{2\pi N_{I}}{\Omega\hbar}\sum_{q}\mid U_q\mid^2\mid F_{\xi,\xi'}(\gamma)\mid^2\delta(E_{\xi}-E_{\xi'})\delta_{k_y,k_y'+q_y}.
 \end{equation}
Here, $N_{I}$ is the impurity density and $\gamma=q^2l_c^2/2$. 
The 2D Fourier transformation of the screened charged impurity potential $U(r)=e^{-k_0r}/{4\pi \epsilon_0\epsilon_r r}$ is
$ U_q=U_0[q^2+k_0^2]^{-1/2}\simeq U_0/k_0$ for short range delta function-like potential,
where $U_0=e^2/4\pi\epsilon_0\epsilon_r$; $k_0$ is the screening vector. And the form factor 
$F_{\xi,\xi'}(\gamma)=<\xi\mid e^{i\vec{q}.\vec{r}}\mid \xi'>$ which can be evaluated for 
elastic scattering, the dominant contribution,  i.e., $n=n'$ as
\begin{equation}
 \mid F_{n,n}(\gamma)\mid^2=e^{-\gamma}[\mid A_{n,s}^{\tau p}\mid^2 L_{n}(\gamma)+\mid B_{n,s}^{\tau p}\mid^2 L_{n-1}(\gamma)]^2.
\end{equation}
Here, $L_{n}(\gamma)$ is the Laguerre polynomial of order $n$.
To proceed further, we replace summation over $k_y$ by $\Omega/(2\pi l_c^2)$,
 $ \sum_{q}\rightarrow\frac{\Omega}{(2\pi)^2}\int q dq d\phi$ and $(x_\xi-x_{\xi'})^2=q_y^2l_c^4=[q\sin\phi]^2l_c^4$ ,
we obtain
\begin{equation}\label{cond}
 \sigma_{xx}\simeq \frac{e^2}{h}\frac{N_{I}U_0^2}{l_c^2k_0^2k_BT}\sum_{n,\tau,s,p}G_{\tau,s}I_{n,s}^{\tau, p}f_n^{\tau,s}(1-f_n^{\tau,s})
\end{equation}
where $I_{n,s}^{\tau, p}=\mid A_{n,s}^{\tau,p}\mid^4(2n+1)+\mid B_{n,s}^{\tau,p}\mid^4(2n-1)-2n\mid A_{n,s}^{\tau,p}\mid^2\mid B_{n,s}^{\tau,p}\mid^2$
with $A_{n,s}^{\tau,p}$ and $B_{n,s}^{\tau,p}$ are given in Eqs(\ref{13}) and (\ref{14}). Also,
$G_{\tau,s}=\Delta_{\tau,s}/(E_g^2+\Delta_{\tau,s}E_{2d}(\alpha+\beta))$. To obtain $I_{n,s}^{\tau,p}$, the following standard integration result 
has been used:
\begin{equation}
 \int_{0}^{\infty}\gamma e^{-\gamma}[L_{n}(\gamma)]^2d\gamma=(2n+1).
\end{equation}
Because of the large momentum separation between two valleys, intervalley scattering is negligibly small.
Spin and valley polarization for longitudinal conductivity can be defined as
\begin{equation}
 P_s=\frac{\sigma_{K}^{\ua}-\sigma_{K}^{\da}+\sigma_{K'}^{\ua}-\sigma_{K'}^{\da}}{\sigma_{K}^{\ua}+\sigma_{K}^{\da}+\sigma_{K'}^{\ua}+\sigma_{K'}^{\da}}
\end{equation}
and
\begin{equation}
 P_v=\frac{\sigma_{K}^{\ua}-\sigma_{K'}^{\ua}+\sigma_{K}^{\da}-\sigma_{K'}^{\da}}{\sigma_{K}^{\ua}+\sigma_{K}^{\da}+\sigma_{K'}^{\ua}+\sigma_{K'}^{\da}}.
\end{equation}
\subsection{ Quantum Hall conductivity}
In linear response regime, the quantum Hall conductivity( $\sigma_{xy}$) is defined as\cite{theory}:
 \begin{equation}\label{qhc}
 \sigma_{xy}=\frac{ie^2\hbar}{\Omega}\sum_{\xi\ne\xi'}[f(E_{\xi})-f(E_{\xi'})]\frac{<\xi\mid \hat{v}_x\mid \xi'><\xi'\mid \hat{v}_y\mid\xi>}
 {(E_{\xi}-E_{\xi'})^2}.
\end{equation}
In the above expression, velocity operators are defined as: $\hat{v}_x=\partial H/\partial p_x$ and $\hat{v}_y=\partial H/\partial p_y$.
It is to be noted that here in (\ref{qhc}), the matrix elements of velocity operators appear which is in general non-zero (see the appendix)
unlike the drift velocity element defined earlier which for our case is zero. 
After performing summation over $k_y$, the above expression simplifies to
\begin{eqnarray}
 \sigma_{xy}&=&\frac{ie^2\hbar}{2\pi l_c^2}\sum_{\zeta\ne\zeta'}[f(E_{\zeta})-f(E_{\zeta'})]
\frac{<\zeta\mid \hat{v}_x\mid \zeta'><\zeta'\mid \hat{v}_y\mid \zeta>}
 {(E_{\zeta}-E_{\zeta'})^2}\nonumber\\
\end{eqnarray}
Note that $\zeta=\{n,s,\tau,p\}$. The velocity operators
are obtained as 
\begin{equation}
\hat{v}_x=(\alpha+\beta\sigma_z)\frac{p_x}{2m_0}+\tau v_F \sigma_x
\end{equation}
and 
\begin{equation}
\hat{v}_y=(\alpha+\beta\sigma_z)\frac{p_y+eBx}{2m_0}+v_F\sigma_y
\end{equation}
 \begin{figure*}
\centering
    \subfigure[]{\includegraphics[width=.32\textwidth,height=20em]{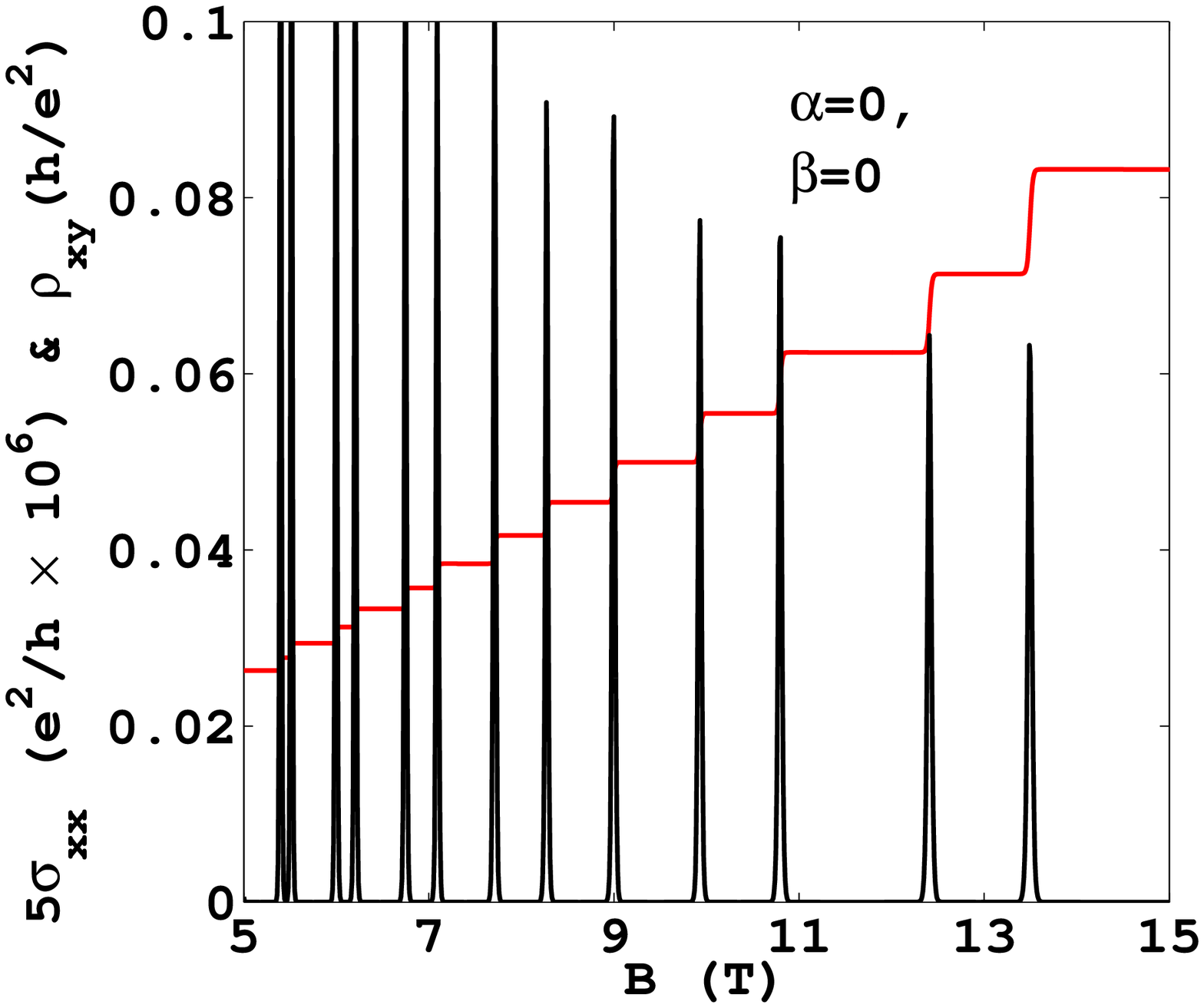}}
   \subfigure[]{\includegraphics[width=.32\textwidth,height=20em]{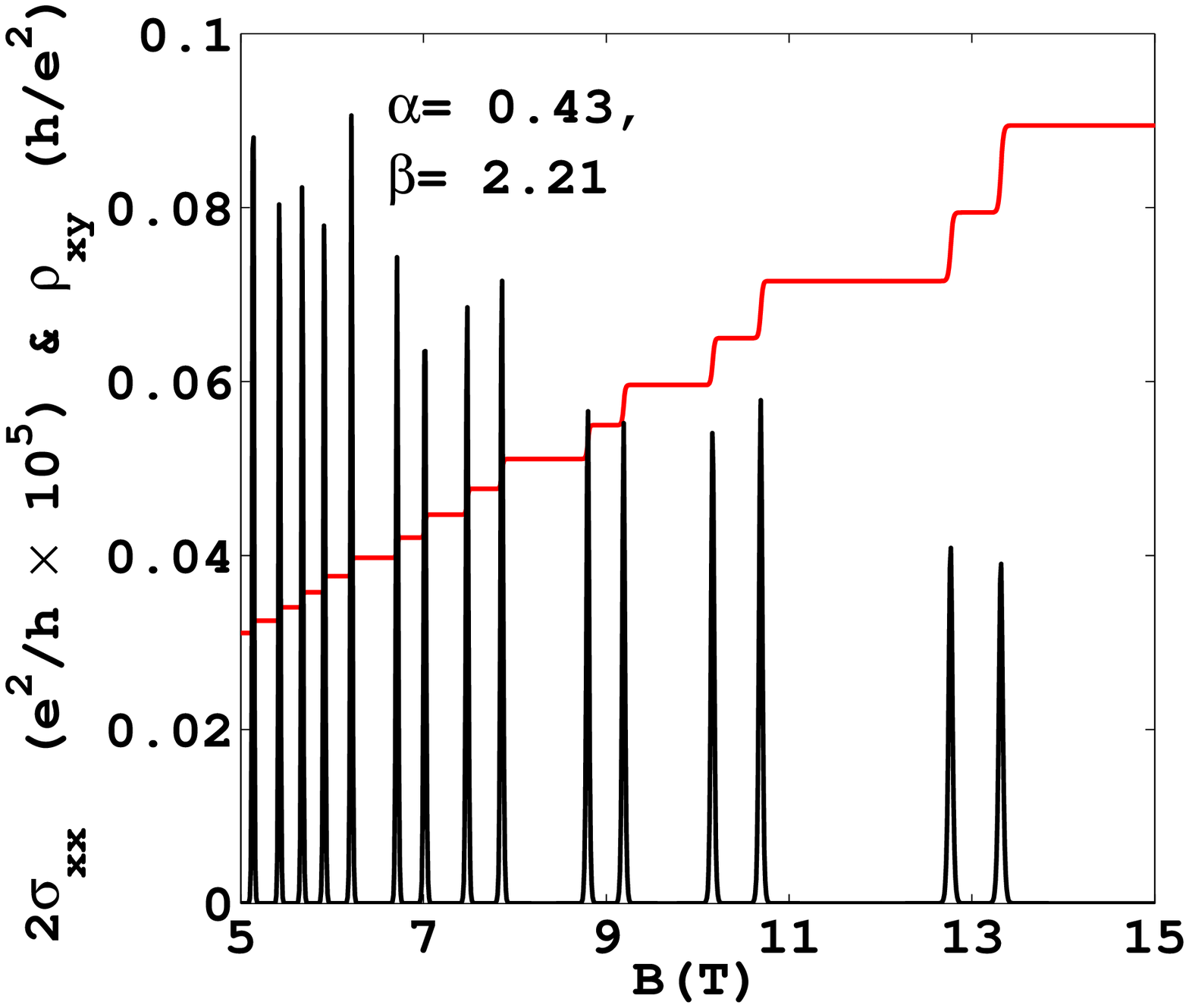}}
   \subfigure[]{\includegraphics[width=.32\textwidth,height=20em]{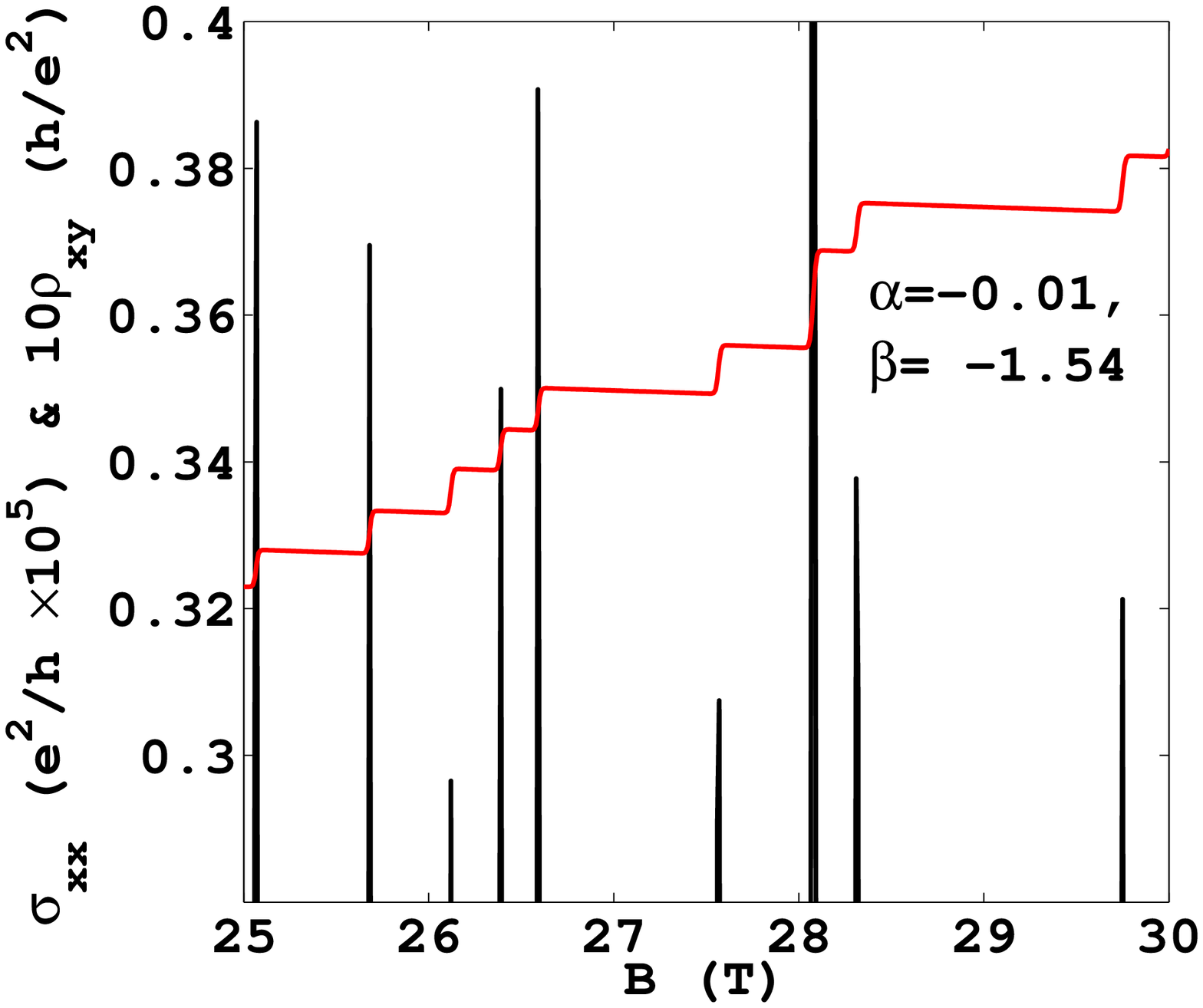}}
   \centering
   \caption{Longitudinal conductivity (black) and Hall resistivity (red) versus magnetic field (B) for different values of 
   topological parameters $\alpha$ and $\beta$. In presence of topological parameters, Landau level spacing is 
   strongly modified which resulted in the appearance of longitudinal conductivity peaks in paired (\ref{Fig7}b) or irregular manner
   (\ref{Fig7}c).}
   \label{Fig7}
 \end{figure*}

Now, as usual the zero level ($n=0$) contribution to the Hall conductivity has to be treated separately as
\begin{eqnarray}
 \sigma^{0,\tau}_{xy}&=&\frac{ie^2\hbar}{2\pi^2l_c^2}\sum_{n',s,p,p'}[f(E_{0,s}^{\tau,p})-f(E_{n',s}^{\tau,p'})]\nonumber\\
&\times&\frac{<0,s,\tau,p\mid \hat{v}_x\mid n',s,\tau,p'><n',s,\tau,p'\mid \hat{v}_y\mid 0,s,\tau,p>}
 {(E_{0,s}^{\tau,p}-E_{n',s}^{\tau,p'})^2}.\nonumber\\
 \end{eqnarray}
 Details of the calculation of the velocity matrix elements are given in appendix.\\
  \begin{figure*}
\centering
    \subfigure[]{\includegraphics[width=.48\textwidth,height=20em]{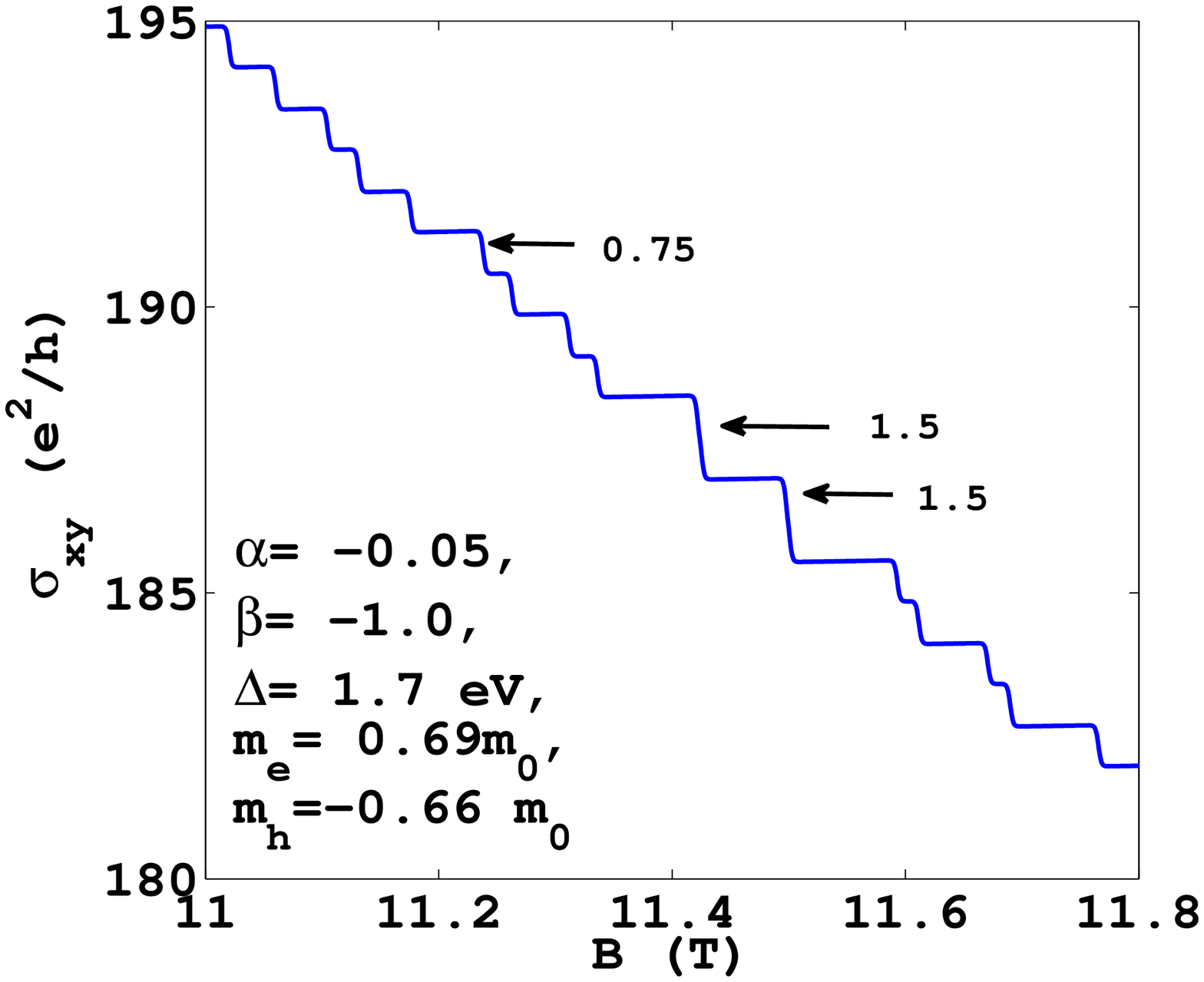}}
   \subfigure[]{\includegraphics[width=.48\textwidth,height=20em]{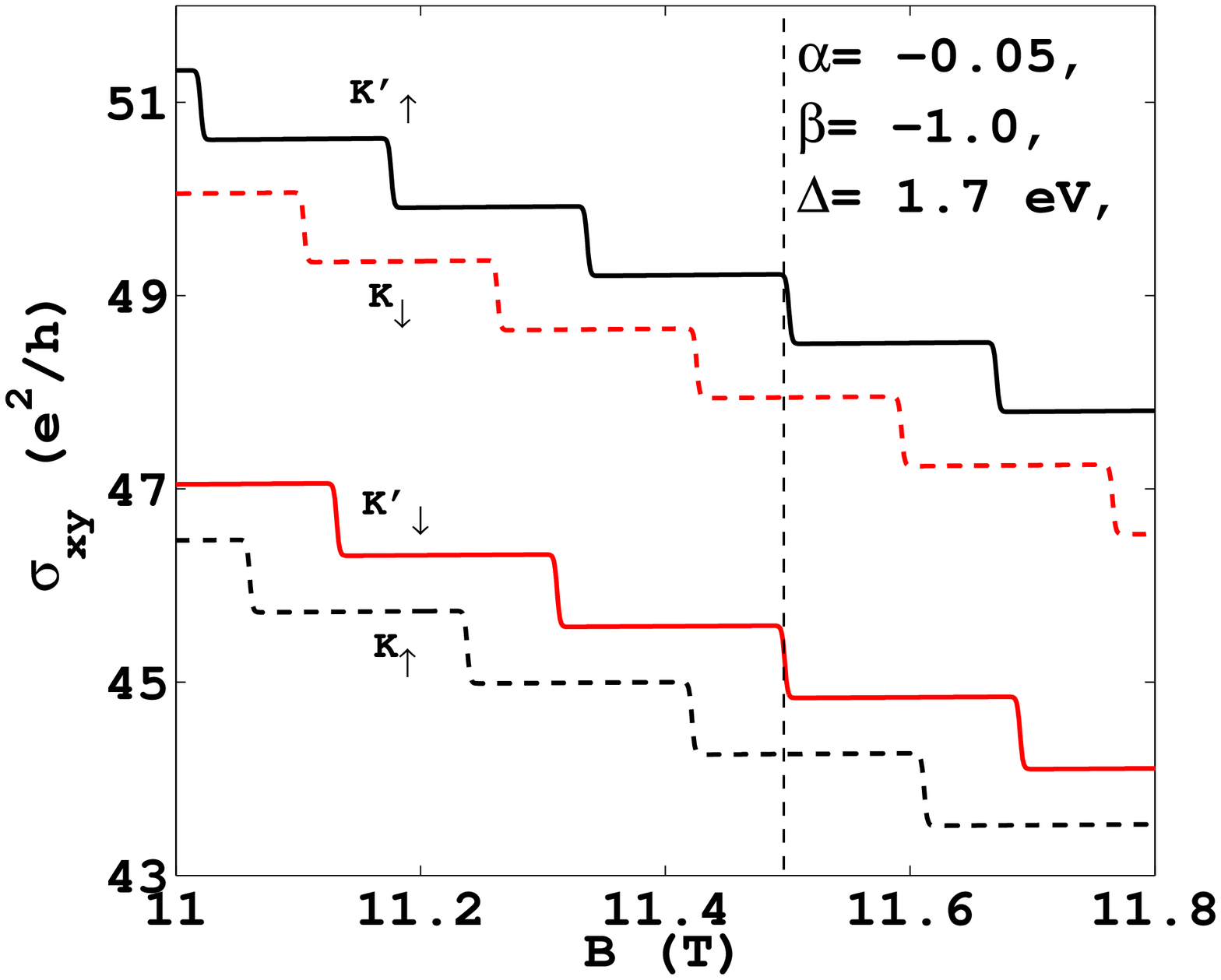}}
     \centering
   \caption{Hall resistivity versus magnetic field (in Tesla). Figure (a) shows the total Hall resistivity while figure (b)
   shows Hall resistivity in each valley. The appearance of big step in (a) is caused by the steps appearing in two valleys
   at the same magnetic field $B=11.5$T as shown in (b).}
   \label{Fig8}
 \end{figure*}
  \begin{figure*}
\centering
  \subfigure[]{\includegraphics[width=.33\textwidth,height=20em]{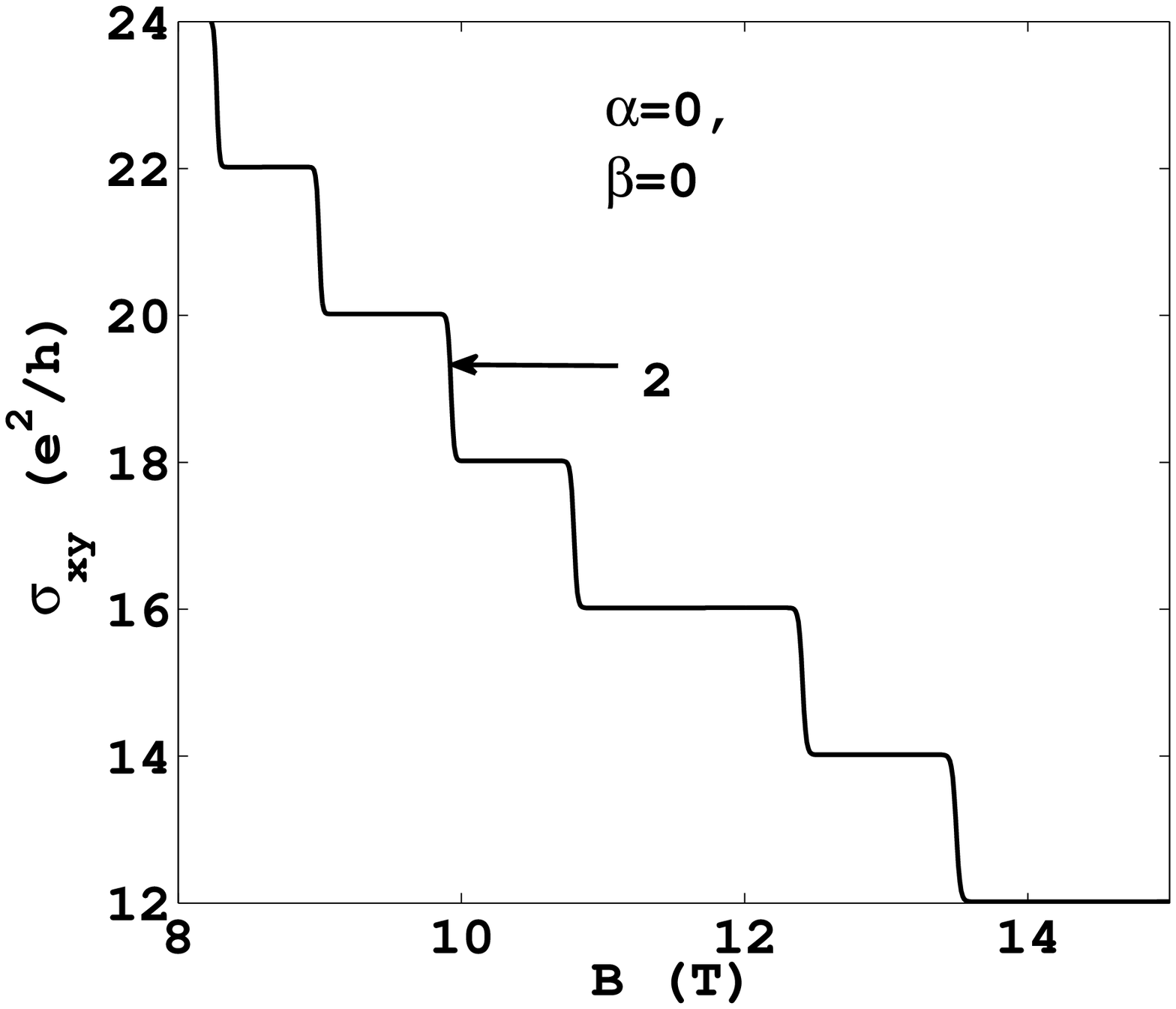}}
   \subfigure[]{\includegraphics[width=.33\textwidth,height=20em]{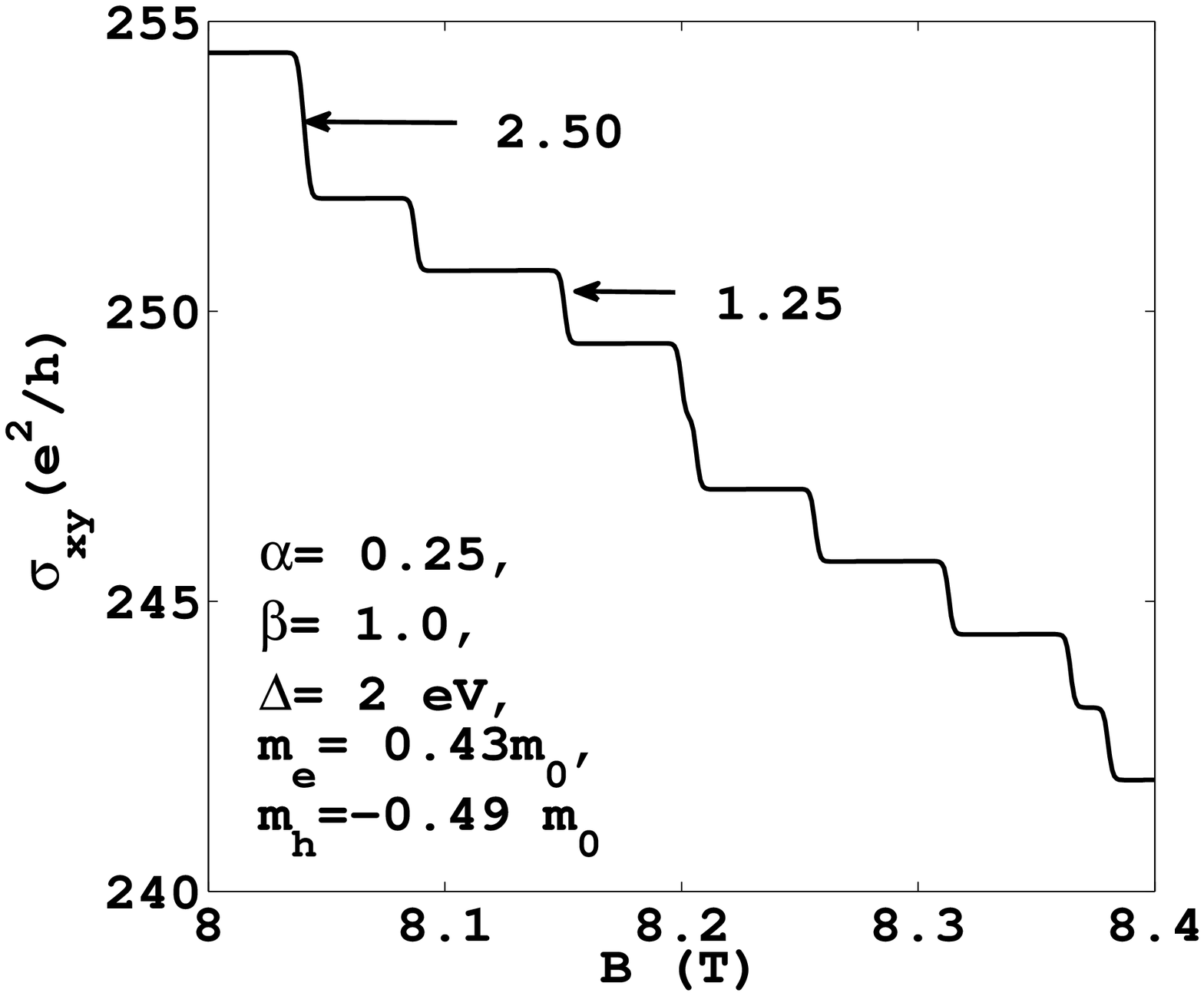}}
   \subfigure[]{\includegraphics[width=.33\textwidth,height=20em]{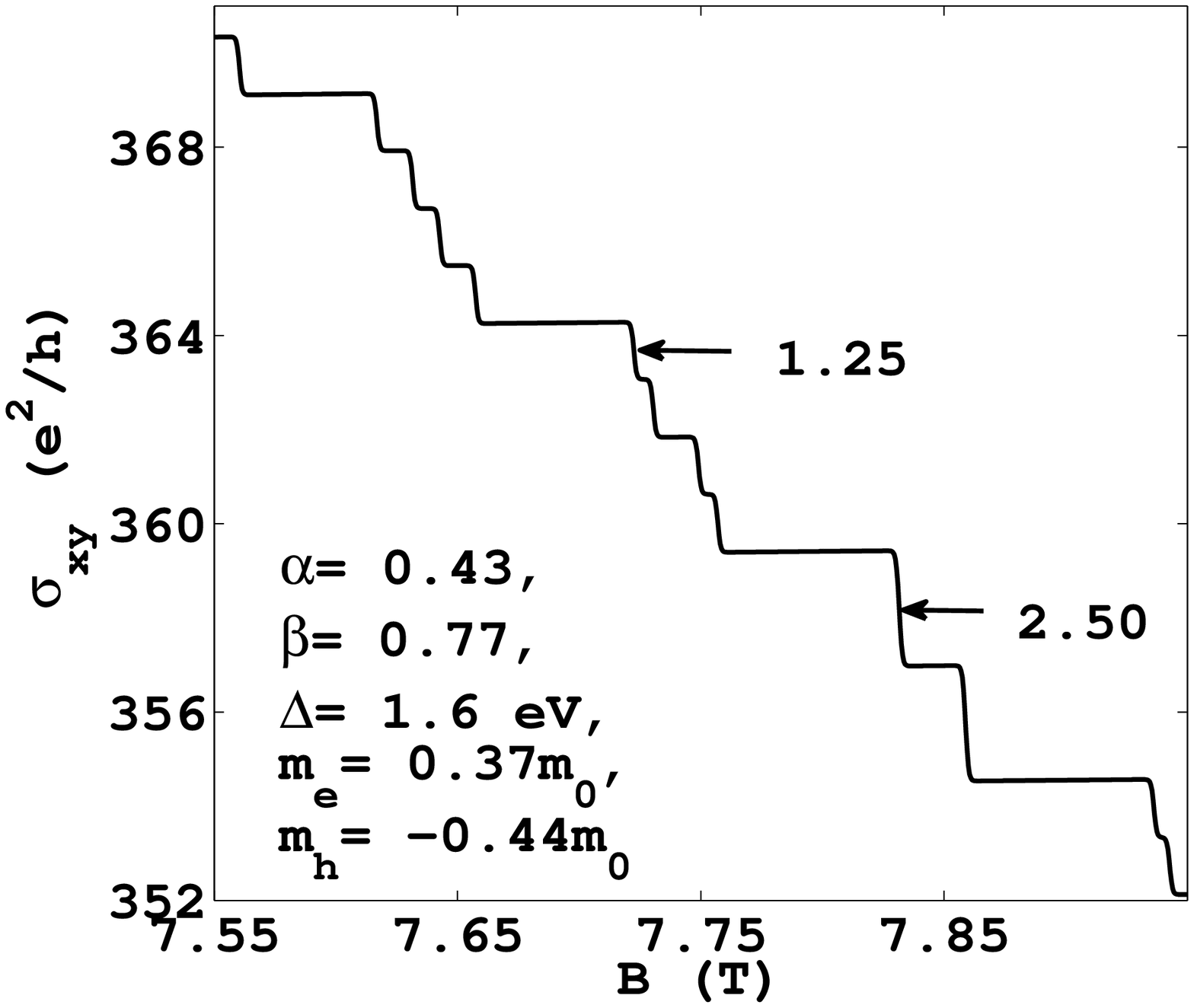}}
   \centering
   \caption{Hall resistivity versus magnetic field (in Tesla) for different values of topological parameters which can be achieved
   by tuning gate voltage. It is shown that not only Hall plateaus but step sizes also get modified. The amplitude of
   Hall resistivity is also increased in presence of topological parameters which can be traced to the effects of 
   topological parameters on Landau levels and velocity matrix elements.}
   \label{Fig9}
 \end{figure*}
\section{Results and discussion}
For numerical plots of longitudinal conductivity and Hall conductivity, we use the following parameters:
Fermi level is at $E_F=0.96$ eV in the conduction band. As gate voltage 
can tune the band gap, we  choose three sets of parameters\cite{reja,reja2}:
$\alpha=0.43$ and $\beta=2.21$ corresponding to $\Delta=1.9$ eV,
$\lambda=0.08$ eV, effective mass of electron $m_e=0.37m_0$ and hole $m_h=-0.44m_0$;
$\alpha=-0.01$ and $\beta=-1.54$ corresponding to $\Delta=1.83$ eV,
$\lambda=0.08$ eV, effective mass of electron $m_e=0.95m_0$ and hole $m_h=-0.94m_0$;
and $\alpha=\beta=0$.
Temperature $(T)= 0.3$ K, impurity density $(N_I)=10^{13} m^{-2}$, screening wave vector:
$k_0=10^7 m^{-1}$, relative permitivity of $MoS_2$: $\epsilon_r=7.3$\cite{mos2_exp,radisa}.

In Figure (\ref{Fig5}), we plot longitudinal conductivity ($\sigma_{xx}$) versus inverse magnetic field (1/B) by using Eq.(\ref{cond}),
to observe SdH oscillations. Figure(\ref{Fig5}a) contains two plots, black is for 
$\alpha=\beta=0$  and red is for non-zero and $\alpha=0.43$ and $\beta=2.21$. 
The black and red plots show appearance of beating pattern in SdH oscillations.
However, the number of beating nodes decreases and oscillations amplitude is damped because of the topological parameter ($\beta=2.21$).
The appearance of SdH oscillations in longitudinal conductivity is the direct consequence of
oscillations in DOS. Total longitudinal conductivity produces beating pattern in SdH oscillation because of
the small difference in frequencies of each spin branch, see Eq. (\ref{ll}) where $s=\uparrow\downarrow$
makes a significance difference in the energy. The beating pattern does not depend on
topological parameter `$\beta$'. However `$\beta$' can influence the number of beating nodes,
this is because of the association of Landau level index `$n$' with $\beta$ in energy spectrum (see Eq.(\ref{ll})).
The origin of the damping in SdH oscillations can be traced to the  behavior of DOS, as shown in the upper panel of Figure(\ref{Fig4}),
which shows two valleys in almost opposite phase because of `$\beta$'.
SdH oscillations for negative value of topological parameter $\beta=-1.54$ and $\alpha=-0.01$ is shown in Figure(\ref{Fig5}b),
where the number of beating nodes increases within a small range of inverse magnetic field.
This is because the Landau levels spacing has become more smaller, as a result within small range of 
magnetic field many Landau levels can pass through Fermi level and thus increase the frequency of the
SdH oscillations. However, the SdH oscillations are enhanced in comparison to without topological parameters.
We conclude that SdH oscillations are damped or pronounced depending upon the sign of topological parameters.
\\
Next we plot spin and valley polarization in longitudinal conductivity versus magnetic field in Figure ({\ref{Fig6}a,b).
We find that 100\% spin polarization can be achieved for finite value of topological parameter as shown by black solid line.
In absence of topological parameter, there is no spin polarization as shown by red solid line. In Figure(\ref{Fig6}b),
we show fully valley polarization can also be achieved.  In compare to Ref.[\onlinecite{tahir_mos2}], fully spin and valley
polarized conductivity are achieved even at low range of magnetic field.
\\
Quantum Hall resistivity ($\rho_{xy}=1/\sigma_{xy}$) and longitudinal conductivity are plotted versus magnetic field in Figure(\ref{Fig7})
by using the formula(\ref{qhc}) and (\ref{cond}). Without topological parameter, plateaus and steps
are increasing slowly with magnetic field in Hall resistivity and sharp peaks of longitudinal conductivity arise at each step of
the Hall resistivity as shown in Figure(\ref{Fig7}a).
The longitudinal conductivity shows peaks at each step of the Hall resistivity,
as shown in Figure (\ref{Fig7}), corresponding to passing of Landau levels through Fermi energy.\\
When $\alpha=0.49$ and $\beta=2.21$, We observe that behavior of steps remain unchanged but longitudinal
conductivity peaks appear in pairs, two nearest pairs are well separated because of the topological parameter induced valley separation.
These SdH peaks appearing in pairs actually correspond to spin-splitting of Landau levels. Small plateaus arise between spin-split
conductivity peaks while longer plateaus arise between two nearest pairs as shown in Figure (\ref{Fig7}b).\\
In the Figure(\ref{Fig7}c), we plot the same for $\alpha=-0.01$ and $\beta=-1.54$, where we see that plateaus
are random in size and a big step arises around $B=28$T. Here, we have shifted the x-axis for better visualization,
as step size becomes too small to be observed properly.\\
To understand the origin of big steps, we plot Hall conductivity
versus magnetic field for different sets of parameters in Figure(\ref{Fig8}).
In Figure(\ref{Fig8}a), we found that steps are not fixed in size i.e.,
there are two different size of steps: $n_{eff}(e^2/h)$ and $(n_{eff}/2)(e^2/h)$ with $n_{eff}=1.5$, indicated by black arrow.
The value of $n_{eff}$ is obtained form numerical datas of Hall conductivity.
The origin of big steps can be explained from Figure(\ref{Fig8}b) where
each spin component in each valley is plotted separately. It shows that
big step arises when spin-up and down components of Hall conductivity of K/K'-valley exhibit a step at the same magnetic field,
as shown by a vertical dashed line.
This phenomena is the consequence of vanishing spin-splitting Landau level in valley space, as discussed in the Landau level plot
(see Figure(\ref{Fig2}b)). For $\alpha=\beta=0$, steps are always $2(e^2/h)$
including spin-splitting as shown in Figure(\ref{Fig9}a).
The enhancement of the quantum Hall conductivity can be traced to the $\alpha$, $\beta$ dependent additional terms which appear in the
velocity matrix elements. Similar patterns are observed for other two sets of parameters (given in the figures) in Figure(\ref{Fig9}b and \ref{Fig9}c).
In Figure (\ref{Fig9}b), steps are found as $2.5 (e^2/h)$ and $1.25(e^2/h)$; same steps appear in Figure (\ref{Fig9}c). 
It should be noted that in all these figures, Hall conductivity steps are always of two types, $n_{eff}$ and 
$n_{eff}/2$ where $n_{eff}$ is the traditional integer quantum Hall step while the $n_{eff}/2$ is the topologically induced
fractional step. Finally we mention that similar phenomena is also found in presence of spin and valley dependent
Zeeman terms\cite{tahir_mos2}, but in our case without Zeeman terms we can also get 
two types of steps which can be controlled by tuning topological parameters
via gate voltage. We conclude that there could be two different origin for additional steps:
one is spin and valley dependent Zeeman terms and another is gate voltage induced topological terms.

\section{Conclusion}
We have studied quantum magneto-transport properties of monolayer $MoS_2$ including the gate voltage controlled topological parameters,
$\alpha$, $\beta$. We found that magnetoconductivity oscillations are strongly affected by these topological parameters. 
When topological parameters are positive there is a suppression of SdH oscillations because of the almost opposite
phase between the oscillations arising from two valleys. When topological parameters are negative, this effect is much stronger
which causes enhancement of SdH oscillations. Beating nodes are decreased and increased for positive and negative values
of topological parameters, respectively. However, beating pattern appears only in the low range of magnetic field
for positive values of topological parameters. Beating pattern is caused by the superposition of two closely
spaced frequencies of two spin branches. Topological parameters does not play any role in beating pattern, it only
induces a phase factor between two spin branches and modify frequencies of SdH oscillations.
The topological parameters '$\beta$' causes a complete separation between two valleys
without any spin and valley dependent Zeeman term, as a result we get fully spin and valley
polarized magnetoconductivity even at low range of magnetic field, in contrast to the case of Ref.[\onlinecite{tahir_mos2}].
In integer quantum Hall effect, fractional Hall steps appear in addition to the usual
integer Hall steps. Quantum Hall steps size  can be tuned by changing topological parameters via gate voltage. The present
study can also be useful for further theoretical works on gate voltage controlled magneto-thermal
properties.
\section{Acknowledgement}
This work is financially supported by the Department of Science and Technology (Nano-
mission), Govt. of India for funds under Grant No. SR/NM/NS-1101/2011.
\section{Appendix}
The velocity matrix elements are calculated for K-valley ($\tau=+$) and $n\ge 1$ as:
\begin{eqnarray}
 <&n&,s,+,p\mid\hat{v}_x\mid n',s,+,p'>\nonumber\\&=&(\alpha+\beta)v_0A_{n,s}^{+,p}A_{n',s}^{+,p'}
 [\sqrt{n'}\delta_{n-1,n'}+\sqrt{n'-1}\delta_{n-1,n'-2}]\nonumber\\&+&
 v_F[A_{n,s}^{+,p}B_{n',s}^{+,p'}\delta_{n-1,n'}+A_{n',s}^{+,p'}B_{n,s}^{+,p}\delta_{n,n'-1}]\nonumber\\
 &+&v_0(\alpha-\beta)B_{n,s}^{+,p}B_{n',s}^{+,p'}[\sqrt{n'+1}\delta_{n,n'+1}+\sqrt{n'}\delta_{n,n'-1}],
\end{eqnarray}
where $v_0=\sqrt{\hbar\omega_c/8m_0}$. Similarly for K'-valley ($\tau=-$) and $n\ge 1$
\begin{eqnarray}
 <&n&,s,-,p\mid\hat{v}_x\mid n',s,-,p'>\nonumber\\&=&(\alpha+\beta)v_0A_{n,s}^{-,p}A_{n',s}^{-,p'}
 [\sqrt{n'+1}\delta_{n,n'+1}+\sqrt{n'}\delta_{n,n'-1}]\nonumber\\&-&
 v_F[A_{n',s}^{-,p}B_{n,s}^{-,p'}\delta_{n,n'-1}+A_{n,s}^{-,p'}B_{n',s}^{-,p}\delta_{n',n-1}]\nonumber\\
 &+&v_0(\alpha-\beta)B_{n,s}^{+,p}B_{n',s}^{+,p'}[\sqrt{n'}\delta_{n-1,n'}+\sqrt{n'-1}\delta_{n-1,n'-2}].\nonumber\\
\end{eqnarray}
The matrix elements of $\hat{v}_y$ are evaluated for $\tau=+1$ and $n\ge1$ as
\begin{eqnarray}
  <&n'&,s,+,p\mid\hat{v}_y\mid n,s,+,p'>\nonumber\\&=&(-i)\Big[(\alpha+\beta)v_0A_{n,s}^{+,p}A_{n',s}^{+,p'}
 \{\sqrt{n'}\delta_{n'-1,n}-\sqrt{n-1}\delta_{n'-1,n-2}\}\nonumber\\&+&
 v_F\{A_{n',s}^{+,p}B_{n,s}^{+,p'}\delta_{n'-1,n}-A_{n,s}^{+,p'}B_{n',s}^{+,p}\delta_{n',n-1}\}\nonumber\\
 &+&v_0(\alpha-\beta)B_{n,s}^{+,p}B_{n',s}^{+,p'}\{\sqrt{n+1}\delta_{n',n+1}-\sqrt{n'}\delta_{n',n-1}\}\Big],
\end{eqnarray}
and for $\tau=-1$ and $n\ge1$ as
\begin{eqnarray}
 <&n'&,s,-,p\mid\hat{v}_y\mid n,s,-,p'>\nonumber\\&=&(-i)\Big[(\alpha+\beta)v_0A_{n,s}^{-,p}A_{n',s}^{-,p'}
 [\sqrt{n'+1}\delta_{n',n+1}-\sqrt{n}\delta_{n',n-1}]\nonumber\\&+&
 v_F\{A_{n',s}^{-,p}B_{n,s}^{-,p'}\delta_{n',n-1}-A_{n,s}^{-,p'}B_{n',s}^{-,p}\delta_{n'-1,n}\}\nonumber\\
 &+&v_0(\alpha-\beta)B_{n,s}^{+,p}B_{n',s}^{+,p'}[\sqrt{n}\delta_{n'-1,n}-\sqrt{n-1}\delta_{n'-1,n-2}]\Big].
 \end{eqnarray}
The velocity matrix elements corresponding to ground states for K-valley ($\tau=+1$) are 
\begin{eqnarray}
 <0,s,+,p\mid \hat{v}_x\mid n',s,+,p'>&=&A_{n',s}^{+,p'}v_F\delta_{0,n'-1}+v_0B_{n',s}^{+,p'}(\alpha-\beta)\nonumber\\
 &\times&[\sqrt{n'+1}\delta_{0,n'+1} +\sqrt{n'}\delta_{0,n'-1}]\nonumber\\
\end{eqnarray}
and
\begin{eqnarray}
 <n',s,+,p\mid \hat{v}_y\mid 0,s,+,p'>&=&(-i)\big[A_{n',s}^{+,p'}v_F\delta_{n'-1,0}\nonumber\\&+&v_0B_{n',s}^{+,p'}(\alpha-\beta)\delta_{n',1}].
\end{eqnarray}
Similarly for K'-valley ($\tau=-1$)
\begin{eqnarray}
 <0,s,-,p\mid \hat{v}_x\mid n',s,-,p'>&=&A_{n',s}^{+,p'}v_0[\delta_{0,n'+1}+\sqrt{n'}\delta_{0,n'-1}]\nonumber\\&-&
 v_FB_{n',s}^{+,p'}(\alpha-\beta)\delta_{0,n'-1}\nonumber\\
\end{eqnarray}
Note that velocity matrix elements are non-zero only for $n'=n\pm1$.


\begin{thebibliography}{99}
 \bibitem{novo} K. S. Novoselov, A. K. Geim, S. Morozov, D. Jiang, Y. Zhang,
                S. Dubonos, I. Grigorieva, and A. A. Firsov, Science 306, 666 (2004).
 \bibitem{neto} A. H. Castro Neto, F. Guinea, N. M. R. Peres, K. S. Novoselov,
                and A. K. Geim, Rev. Mod. Phys. 81, 109 (2009).
 \bibitem{park} K. F. Mac, K. F. McGill, J. Park and P. L. McEuen, Science 344, 1489 (2014).
 \bibitem{carbotte} Z. Li and J. P Carbotte, Phys. Rev. B 86, 205425 (2012).
 \bibitem{tahir1} M. Tahir, A. Manchon and U. Schwingenschlogl, Phys. Rev. B 90, 125438 (2014)
 \bibitem{xiao} D. Xiao, G. Liu, W. Feng, X. Xu and W. Yao, Phys. Rev. Lett. 108, 196802 (2012)
 \bibitem{bandgap} K. F. Mak, C. Lee, J. Hone, J. Shan, and T. F. Heinz, Phys. Rev.Lett. 105, 136805 (2010);
 \bibitem{photo1} A. Splendiani, L. Sun, Y. Zhang, T. Li, J. Kim, C. Y. Chim, G. Galli, and F. Wang, Nano Lett. 10, 1271 (2010);
 \bibitem{photo2} T. Korn, et. al., Appl. Phys. Lett. 99, 102109 (2011).
 \bibitem{photo3} K. F. Mak, K. He, J. Shan, and T. F. Heinz, Nat. Nanotechnol. 7, 494 (2012)
 \bibitem{photo4} H. Zeng, J. Dai, W. Yao, D. Xiao, and X. Cui, Nat. Nanotechnol. 7, 490 (2012)
 \bibitem{transistor} B. Radisavljevic, A. Radenovic, J. Brivio, V. Giacometti, and A. Kis, Nat. Nanotechnol. 6, 147 (2011).
 \bibitem{loss} Jelena Klinovaja and Daniel Loss, Phys. Rev. B 88, 075404 (2013).
 \bibitem{valley2} S. Wu, J. S. Ross, G.-B. Liu, G. Aivazian, A. Jones, Z. Fei, W.
                    Zhu, D. Xiao, W. Yao, D. Cobden, and X. Xu, Nat. Phys. 9, 149 (2013).
 \bibitem{graphene1} V. P. Gusynin and S. G. Sharapov, Phys. Rev. Lett. 95, 146801 (2005).
 \bibitem{tahir_gra2} M. Tahir and K. Sabeeh, J. Phys.: Condens. Matter 24, 135005 (2012).
 \bibitem{vasilo_gra} P. M. Krstajic and P. Vasilopoulos, Phys. Rev. B 86, 115432 (2012).
 \bibitem{gra_exp1} Yuanbo Zhang, Yan-Wen Tan, Horst L. Stormer and Philip Kim, Nature 438, 201 (2005)
 \bibitem{gra_exp2} K. S. Novoselov, Z. Jiang, Y. Zhang, S. V. Morozov,
                    H. L. Stormer, U. Zeitler, J. C. Maan, G. S. Boebinger, P. Kim, A. K. Geim, Science 315, 1389 (2007).
 \bibitem{tahir_sili} M. Tahir and U. Schwingenschlogl, Sci. Rep. 3, 1075 (2013).
 \bibitem{vasilo_sili} Kh. Shakouri, P. Vasilopoulos, V. Vargiamidis, and F. M. Peeters,
                       Phys. Rev. B 90, 235423 (2014).
 \bibitem{mos2_exp} X. Cui, G.-H. Lee, Y. D. Kim, G. Arefe, P. Y. Huang,
                    C.-H. Lee, D. A. Chenet, X. Zhang, L. Wang, F. Ye, F. Pizzocchero, B. S. Jessen, K. Watanabe, T. Taniguchi, D. A.
                    Muller, T. Low, P. Kim, and J. Hone, Nat. Nanotechnol. 10, 534 (2015)
 
 \bibitem{tahir_mos2} M. Tahir, P. Vasilopoulos and F. M. Peeters, Phys. Rev. B 93, 035406 (2016).
 \bibitem{fan} X. Li, F. Zhang, and Q. Niu, Phys. Rev. Lett. 110, 066803 (2013).
 \bibitem{fan_exp} Z. Wu, S. Xu, H. Lu, G. Liu, A. Khamoshi,
                   T. Han, Y. Wu, J. Lin, G. Long, Y. He, Y. Cai, F. Zhang, N. Wang, arxiv 1511:00077.
 \bibitem{reja} Habib Rostami and Reja Asgari, Phys. Rev. B 91, 075433 (2015).
 \bibitem{reja2} Habib Rostami, Ali G. Moghaddam, and Reza Asgari, Phys. Rev. B 88, 085440 (2013).
 \bibitem{ab_initio} S. Lebegue and O. Eriksson, Phys. Rev. B 79, 115409 (2009).
 \bibitem{burkard} A. Kormanyos, V. Zolyomi, N. D. Drummond, P. Rakyta, G. Burkard, and V. I. Falko, Phys. Rev. B 88, 045416 (2013).
 \bibitem{felix} Felix Rose, M. O. Goerbig, and Frederic Piechon, Phys. Rev. B 88, 125438 (2018).
 \bibitem{vasilo_rashba} X. F. Wang and P. Vasilopoulos, Phys. Rev. B 67, 085313 (2003).
 \bibitem{broadening} Y. Zheng and T. Ando, Phys. Rev. B 65, 245420 (2002);
                      T. Stauber, N. M. R. Peres, and F. Guinea, ibid. 76, 205423  (2007).
 \bibitem{peeters_92} F. M. Peeters and P. Vasilopoulos, Phys. Rev. B 46, 4667 (1992).
 \bibitem{theory} M. Charbonneau, K. M. Van Vliet, and P. Vasilopoulos, J. Math. Phys. 23, 318 (1982).
 \bibitem{radisa} B. Radisavljevic and A. Kis, Nat. Mater. 12, 815 (2013).
 \end{thebibliography}
\end{document}